\documentclass[aps,pra,amssymb,amsmath,%
    twocolumn,twoside,%
    showpacs,noshowkeys]{revtex4}

\usepackage{graphicx}

\newcommand*{\ket}[1]{|{#1}\rangle}
\newcommand*{\bra}[1]{\langle{#1}|}
\newcommand*{\braket}[2]{\langle{#1}|{#2}\rangle}
\renewcommand*{\Re}{\mathrm{Re}\,}
\renewcommand*{\Im}{\mathrm{Im}\,}

\DeclareMathOperator{\diag}{diag}

\begin{document}
\title{Multilevel Holstein-Primakoff approximation and its
  application to atomic spin squeezing and ensemble quantum memories}
\author{Z. Kurucz}
\author{K. M{\o}lmer}
\affiliation{Lundbeck Foundation Theoretical Center for Quantum System Research, Department of Physics and Astronomy, University of Aarhus, 8000 \r Arhus C, Denmark}


\date{Dec.\ 20, 2009}
\begin{abstract}
  We show that an ensemble of identical $d$-level atoms can be efficiently described by $d-1$ collective oscillator degrees of freedom in the vicinity of a product state with all atoms in the same, but otherwise arbitrary single-particle state.  We apply our description to two different kinds of spin squeezing: (i) when each spin-$F$ atom is individually squeezed without creating interatomic entanglement and (ii) when a particular collective atomic oscillator mode is squeezed via quantum non-demolition (QND) measurement and feedback.  When combined in sequence, the order of the two methods is relevant in the final degree of squeezing.  We also discuss the role of the two kinds of squeezing when multi-sublevel atoms are used as quantum memories for light.
\end{abstract}
\maketitle

\section{Introduction}
\label{sec:introduction}

Large ensembles of identical particles are excellent candidates for light-matter interfaces, quantum memories, repeaters for long distance quantum communication, and registers for quantum computing.  This is not only due to the fact that in these systems certain collective quantum degrees of freedom efficiently couple to the electromagnetic radiation field, but also because the desired control is provided by simple, experimentally accessible  interaction mechanisms.  This is the case, e.g., for large collections of $N$ identical spin-$\frac12$ particles.  Such an ensemble can be effectively described using a single collective spin, if the initial state is a pure product state with all particles prepared in the same single-particle state, and if the Hamiltonian describing the system is a sum of identical single-particle operators.  Since, in this case, the quantum state of the entire collection remains invariant under permutations of particles, the corresponding restricted Hilbert subspace is equivalent to the states of a large spin-$\tfrac N2$ spin.  If, in addition, the dynamics only weakly perturbs the initial state, the collective states explore only a limited range of states, and the pseudospin eigenbasis can be mapped onto an oscillator basis of states, and one may benefit from the simple and well studied properties of harmonic oscillator systems.

In this paper we wish to generalize the oscillator description of two-level systems to multilevel particles.  A natural example is the one of atoms having a Zeeman degenerate ground state with total angular momentum $F$ or a hyperfine ground state manifold with a range of angular momentum quantum numbers $F'$ and associated $|M'| \leq F'$.  In a large ensemble of such atoms, the collective spin picture is less useful, and we will proceed directly to an effective oscillator description of the symmetric collective states of the system and of the system dynamics. Various methods allow the control of single particle states, and the atoms in a large ensemble can be prepared in essentially any superposition of the hyperfine ground states~\citep{PhysRevLett.99.163002}.  Therefore, it is a particular purpose of our work to develop a theory of collective states in the vicinity of a product state of arbitrary single atom states, and to determine the interplay between the collective and single particle properties of such samples.

In Sec.~\ref{sec:HP-gen}, we review the usual Holstein-Primakoff approximation, describing a collection of spin-$\frac12$ particles by a single harmonic oscillator degree of freedom, and we generalize this description to multilevel systems expanded around product states of arbitrary single particle state vectors.  We discuss the representation of collective operators, which are a sum of single particle operators over the entire ensemble, and we demonstrate that in the vicinity of product states, such operators can be associated with oscillator quadrature operators in a generalized Holstein-Primakoff approximation. In Sec.~\ref{sec:spsq}, we turn to a special discussion of spin squeezing, and we point out the formal distinction between the effect of spin squeezing within each $F > \tfrac12$ hyperfine angular momentum manifold and squeezing of collective spin degrees of freedom.  We also show that, when applied sequentially, the order of the two kinds of squeezing is relevant.  In Sec.~\ref{sec:qmem}, we discuss the use of samples of multi-sublevel atoms as quantum memories for light and the role of internal spin squeezing in such atoms. Sec.~\ref{sec:conclusions} concludes the paper.


\section{Multilevel Holstein-Primakoff approximation}
\label{sec:HP-gen}

A convenient way to describe an ensemble of spin-$\frac12$ particles or two-state atoms which are homogeneously coupled to external perturbations is in terms of the collective spin operator $\hat{\mathbf J} = \sum_{j=1}^N \hat{\mathbf s}^{(j)}$. If the initial state of the system is a product state with all members occupying the same pure state, the dynamically accessible Hilbert space is characterized by a single angular momentum ladder.  It is then straightforward to introduce oscillator-like creation and annihilation operators in the Holstein-Primakoff representation,
\begin{gather}
  \label{eq:standard-HP}
  \hat J_{x-} \equiv \hat a^\dag \sqrt{2J-\hat a^\dag \hat a},
  \qquad
  \hat J_x \equiv J- \hat a^\dag \hat a,
\end{gather}
where $J$ is the total angular momentum quantum number, which is a constant of motion in the cases considered, and the $x$ quantization axis is chosen to be the direction of polarization in the initial product state.  For macroscopic polarization ($J\gg1$) and if the system stays in the vicinity of the spin coherent state $\ket{J, m_J^x=J}$, the atomic oscillator picture is especially efficient and it allows us to directly define the quasi-canonical atomic quadrature operators,
\begin{gather}
  \label{eq:standard-XP}
  \hat X \equiv \frac1{\sqrt J} \hat J_y,
  \qquad
  \hat P \equiv \frac1{\sqrt J} \hat J_z,
\end{gather}
with $[\hat X, \hat P] = i\hat J_x / J \approx i$.

For an ensemble consisting of atoms with larger angular momenta $F > \frac12$, it is also possible to define the collective spin operator $\hat{\mathbf J} = \sum_{j=1}^N \hat{\mathbf F}^{(j)}$.  If each atom is initially prepared in the same spin coherent state, and the Hamilton operator is an element of the operator algebra generated by the components of $\hat{\mathbf J}$, this collective spin again provides a general description of the system. In that case, it is possible to think of each atom as a collection of $2F$ fictitious spin-$\frac12$ particles, and to think of the initial state as a symmetric product state of the total of $2FN$ fictitious spin-$\frac12$ particles, which evolves in the manifold of states of the large collective spin with $J=NF$. The collective spin components constituting the Hamiltonian are invariant under the exchange of any two fictitious spin-$\frac12$ particles irrespective of whether they belong to the same atom or to different atoms. The collective spin description is, however, incomplete in general.  For example, the atoms may be individually prepared in arbitrary internal superposition states $\sum_{m} c_m|F,m_F\rangle$ corresponding to correlations (entanglement) among their own fictitious spin-$\frac12$ constituents.  Even if the atoms are thus all prepared in the same state, the $2FN$ fictitious spins are not equivalent: those pertaining to the same atom share correlations in the initial atomic product state, while the ones that pertain to different atoms do not.

We will now introduce a generalization of the Holstein-Primakoff representation that is capable of treating not only the total angular momentum operators but any permutation invariant sum of single-particle operators of an ensemble of $d$-level systems.

\subsection{Generalized spin operators and collective atomic oscillators}
\label{sec:spinops}

Let us consider an ensemble of $N$ identical $d$-level systems, and
let $\{\ket{\phi_\alpha}\}_{\alpha=0}^{d-1}$ denote an arbitrary
orthonormal single-particle basis.  Throughout the paper, we restrict
ourselves to symmetric collective states, i.e., those invariant under
permutation of the internal state of any two particles.  For
spin-$\frac12$ particles, this corresponds to the maximal total
angular momentum manifold, and our representation coincides with the
standard approach.

In the general case, the symmetric subspace is spanned by the
occupation number states
\begin{multline}
  \label{eq:N-body-basis}
  \ket{n_0,m_1,l_2,\ldots} \equiv
  \frac1{\sqrt{n!\, m!\, l! \ldots}}
  \sum_{\text{perm}}
  \ket{\phi_0}_1 \cdots \ket{\phi_0}_n
  \\ \times
  \ket{\phi_1}_{n+1} \cdots \ket{\phi_1}_{n+m}
  \; \ket{\phi_2}_{n+m+1} \cdots,
\end{multline}
which means that $n$ atoms are in the internal state $\ket{\phi_0}$,
$m$ in $\ket{\phi_1}$, $l$ in $\ket{\phi_2}$, etc., and they sum up to
$N = n + m + l + \ldots$ atoms.  It is easy to see that the collective
operators
\begin{gather}
  \label{eq:Sigma}
  \hat \Sigma_{\alpha\beta} \equiv \sum_{j=1}^N
  \ket{\phi_\alpha}_j {}_j\bra{\phi_\beta},
  \qquad
  (\alpha,\beta = 0,1,\ldots,d-1)
\end{gather}
keep the symmetry of the sub-Hilbert space in consideration.
Furthermore, we observe the following properties,
\begin{align}
  \hat \Sigma_{\alpha\beta} \ket{n_\alpha, m_\beta, \ldots}
  &= \sqrt{(n+1)m}
  \nonumber\\&\quad\times
  \ket{(n+1)_\alpha, (m-1)_\beta, \ldots},
  \label{eq:Sigma-step}
  \\
  \hat \Sigma_{\alpha\alpha} \ket{n_\alpha, m_\beta, \ldots}
  &= n \; \ket{n_\alpha, m_\beta, \ldots},
  \label{eq:Sigma-number}
\end{align}
and the commutator identity,
\begin{align}
  \big[ \hat\Sigma_{\alpha\gamma}, \hat\Sigma_{\gamma\beta} \big]
  &= \hat \Sigma_{\alpha\beta}
  - \delta_{\alpha\beta} \hat \Sigma_{\gamma\gamma}.
  \label{eq:Sigma-commute}
\end{align}
All other commutators, which can not be brought on the form
$\big[ \hat\Sigma_{\alpha\gamma}, \hat\Sigma_{\gamma\beta} \big]$, vanish.

Any pure state of a spin-$\frac12$ particle is a spin coherent state, i.e., a spin up state along a suitably chosen axis, say~$x$.  The corresponding ensemble state  $\ket{J, m_J^x=J}$ serves as the natural reference state: it is the ground state of the effective atomic oscillator in the Holstein-Primakoff approximation.  With a general single-atom basis vector $\ket{\phi_0}$, we take the product state $\bigotimes^N \ket{\phi_0}$ as the reference state and introduce $d-1$ independent collective atomic oscillator modes with creation and annihilation operators that redistribute the atomic populations between $\ket{\phi_0}$ and the other basis states $\ket{\phi_\alpha}$,
\begin{gather}
  \hat a_\alpha^\dag
  \ket{n_0, m_\alpha, \ldots}
  \equiv \sqrt{m+1} \;
  \ket{(n-1)_0, (m+1)_\alpha, \ldots},
  \label{eq:a-step}
  \\
  \hat a_\alpha^\dag \hat a_\alpha
  \ket{n_0, m_\alpha, \ldots}
  = m \; \ket{n_0, m_\alpha, \ldots},
  \label{eq:a-number}
\end{gather}
with $\alpha = 1, \ldots, d-1$.  The reference state itself corresponds to the multi-mode vacuum state of the oscillators, while the excitation number states coincide with the symmetric states \eqref{eq:N-body-basis}.  In analogy with the Holstein-Primakoff representation, directly comparing Eq.~\eqref {eq:Sigma-step} with \eqref {eq:a-step} and Eq.~\eqref {eq:Sigma-number} with \eqref {eq:a-number}, respectively, the generalized collective spin operators can be expressed as follows,
\begin{gather}
  \label{eq:Sigma_a0}
  \hat\Sigma_{\alpha\beta} = \hat a_\alpha^\dag \hat a_\beta,
  \qquad
  \hat\Sigma_{\alpha0} = \hat a_\alpha^\dag \sqrt{\textstyle N -
    \sum_{\beta\ne0} \hat a_\beta^\dag \hat a_\beta},
\end{gather}
furthermore, using Eq.~\eqref{eq:Sigma-commute},
\begin{gather}
  \label{eq:Sigma_00}
  \hat\Sigma_{00} =
  N - \sum_{\alpha\ne0} \hat a_\alpha^\dag \hat a_\alpha.
\end{gather}

\subsection{Linearization around the reference state:
  Holstein-Primakoff expansion of collective operators}
\label{sec:linarization}

In the vicinity of the reference state $\bigotimes^N \ket{\phi_0}$, i.e., for small number of excitations of the atomic oscillators, only the population $\hat\Sigma_{00}$ has macroscopic expectation value, and the spin operators can be approximated as
\begin{gather}
  \label{eq:HP-approx}
  \hat\Sigma_{\alpha\beta} = \hat a_\alpha^\dag \hat a_\beta
  \;\; \ll \;\;
  \hat\Sigma_{\alpha0} \approx \sqrt N \hat a_\alpha^\dag
  \;\; \ll \;\;
  \hat\Sigma_{00} \approx N.
\end{gather}
This allows us to simplify collective operators that are permutation invariant sums of single-particle operators, or belong to the algebra generated by such operators.

Since $\{ \ket{\phi_\alpha} \bra{\phi_\beta}\}_{\alpha\beta}$ forms an orthonormal single-particle operator basis, for any Hermitian single-particle operator $\hat{\mathcal O}^{(1)}$ the corresponding $N$-body operator can be expanded as
\begin{gather}
  \label{eq:O-expanded}
  \hat{\mathcal O} \equiv \sum_{j=1}^N \hat{\mathcal O}^{(j)}
  = \sum_{\alpha\beta}
  \mathcal O_{\alpha\beta}
  \hat \Sigma_{\alpha\beta},
\end{gather}
where we used the shorthand $\mathcal O_{\alpha\beta} =
\bra{\phi_\alpha} \hat{\mathcal O}^{(1)} \ket{\phi_\beta}$.  Now we distinguish between two cases.  If $\ket{\phi_0}$ is an eigenstate of $\hat{\mathcal O}^{(1)}$, then $\mathcal O_{\alpha0} = 0$ for all $\alpha\ne0$ and the collective operator becomes a c-number plus a correction that is quadratic in the creation and annihilation operators,
\begin{gather}
  \label{eq:O-approx2}
  \hat{\mathcal O} =  N \mathcal O_{00} + \sum_{\alpha\beta\ne0}
  (\mathcal O_{\alpha\beta} - \mathcal O_{00} \delta_{\alpha\beta})
  \hat a_\alpha^\dag   \hat a_\beta.
\end{gather}
For example, taking the spin projection eigenstates $\ket{\phi_\alpha} = \ket{F, m_F^x = F-\alpha}$, the longitudinal component of the total angular momentum is $\hat J_x = \sum_\alpha (F-\alpha) \hat{\Sigma}_{\alpha\alpha} = NF - \sum_{\alpha\ne0} \alpha \, \hat a_\alpha^\dag \hat a_\alpha$.

If $\ket{\phi_0}$ is not an eigenstate of $\hat{\mathcal O}^{(1)}$, we get cross terms between the $\alpha = 0$ and the $\alpha \neq 0$ components, and we obtain the dominant contribution 
\begin{gather}
  \label{eq:O-approx1}
  \hat{\mathcal O} \approx N \mathcal O_{00} + \sqrt N \sum_{\alpha\ne0}
  \big( \mathcal O_{\alpha0} \hat a_\alpha^\dag +
  \mathcal O_{0\alpha} \hat a_\alpha \big).
\end{gather}
Apart from a constant, Eq.~\eqref {eq:O-approx1} is a linear function
of the creation and annihilation operators, hence, it is a linear
function of the collective atomic quadrature variables
\begin{gather}
  \label{eq:atom-XP}
  \hat X_\alpha \equiv
  (\hat a_\alpha + \hat a_\alpha^\dag) / \sqrt2
  \quad \mbox{and} \quad
  \hat P_\alpha \equiv
  (\hat a_\alpha - \hat a_\alpha^\dag) / i\sqrt2.
\end{gather}
Namely, Eq.~\eqref{eq:O-approx1} can be written as
\begin{gather}
  \label{eq:O-approx1b}
  \hat{\mathcal O} \approx N \mathcal O_{00}
  + \sqrt{2N} \sum_{\alpha\ne0} \big(
  \Re \mathcal O_{\alpha0} \hat X_\alpha +
  \Im \mathcal O_{\alpha0} \hat P_\alpha \big).  
\end{gather}

Let us choose real numbers $\xi_0$ and $\xi_1$, and the single-particle basis vector $\ket{\phi_1}$ such that
\begin{gather}
  \label{eq:phi1}
  \hat{\mathcal O}^{(1)} \ket{\phi_0}
  = \xi_0 \ket{\phi_0}  + \xi_1  \ket{\phi_1}.
\end{gather}
Normalization implies that $\xi_0$ and $\xi_1$ are the mean and
variance of the single-particle operator, respectively,
\begin{align}
  \label{eq:xi0}
  \xi_0 &= \bra{\phi_0} \hat{\mathcal O}^{(1)} \ket{\phi_0}
  \equiv \langle {\mathcal O} ^{(1)} \rangle_0,
  \\
  \label{eq:xi1}
  \xi_1^2 &= \bra{\phi_0} \hat{\mathcal O} ^{(1)}
  {}^2 \ket{\phi_0} - \xi_0^2
  \equiv \big( \Delta {\mathcal O} ^{(1)} \big)^2_0 .
\end{align}
Eq.~\eqref {eq:O-approx1b} now simplifies to $\hat{\mathcal{O}} \approx N \xi_0 + \sqrt {2N} \, \xi_1 \hat X_1$.  We can assign to $\hat{\mathcal O}$ a single oscillator quadrature variable
\begin{gather}
  \label{eq:X-from-O}
  \hat X_1 = \frac{\hat{\mathcal O} - \langle \mathcal O \rangle _0}
  {\sqrt{2 (\Delta \mathcal O)^2_0}},
\end{gather}
and in the reference product state of the system, by construction, this collective atomic oscillator is in the ground state and $(\Delta X_1)^2_0 = \tfrac12$.

As an example, we mention that the natural normalization factor for the quadrature operators \eqref{eq:standard-XP}, that are assigned to the transverse angular momentum components, is not the spin quantum number $J$ nor the macroscopic expectation value of $\hat J_x$.  Rather, it is related to the variance of $\hat J_y$ and $\hat J_z$ in the reference state, respectively,
\begin{gather}
  \hat X_y = \frac {\hat J_y} {\sqrt{2(\Delta J_y)_0^2}},
  \quad
  \hat P_z = \frac {\hat J_z} {\sqrt{2(\Delta J_z)_0^2}}.
\end{gather}
Here $(\Delta J_y)_0^2 = (\Delta J_z)_0^2 = J/2$ only for the coherent spin state, but not for a generic reference state.  As long as the reference state is a product state and there is no entanglement among the particles, the corresponding collective atomic oscillators are in the ground state.  Finally, we note that the Heisenberg uncertainty relation implies $[\hat X_y, \hat P_z]/i \le 1$.  The reason why $\hat X_y$ and $\hat P_z$ may have a non-canonical commutation relation is discussed in the next subsection.

\subsection{Operators acting on different oscillators}
\label{sec:overlap}

Let us now consider two collective operators, $\hat A$ and $\hat B$, that can be linearized in the above manner.  As we have shown, we can assign to them the oscillator quadrature operators $\hat X_A$ and $\hat P_B$, respectively, such that
\begin{align}
  \label{eq:X_A-implicit-def}
  \hat A &= \langle A \rangle_0 +
  \sqrt{2 (\Delta A)^2_0} \hat X_A,
  \\
  \label{eq:X_B-implicit-def}
  \hat B &= \langle B \rangle_0 +
  \sqrt{2 (\Delta B)^2_0} \hat P_B.
\end{align}
Note that the position or momentum quadratures of the oscillator modes are fully equivalent and the distinction  depends on the choice of the single-particle basis and, in particular, the phase of the basis vectors. For the two operators, it remains a question, however, whether we can choose the basis in such a way that $\hat X_A$ and $\hat P_B$ are quadratures conjugate to each other acting on the same oscillator, or whether they belong to two independent atomic oscillators. To answer this question in general, we introduce the unnormalized vectors
\begin{align}
  \label{eq:a-vector}
  \ket a &\equiv
  \big( \hat A^{(1)} - \langle A^{(1)} \rangle_0 \big) \ket{\phi_0},
  \\
  \label{eq:b-vector}
  \ket b &\equiv
  \big( \hat B^{(1)} - \langle B^{(1)} \rangle_0 \big) \ket{\phi_0},
\end{align}
and analyze the following three cases.

\paragraph{Parallel case.}

If $\ket a$ and $\ket b$ are parallel to each other ($\ket b = \lambda e^{i\varphi} \ket a$), then $\hat X_A$ and $\hat P_B$ belong to the same atomic oscillator.  Indeed, we can set $\ket{\phi_1} = \ket a /
\|a\|$, so that
\begin{gather}
  \label{eq:Xab-parallel}
  \hat X_A = \hat X_1,
  \qquad
  \hat P_B = \hat X_1 \cos\varphi + \hat P_1 \sin\varphi,
\end{gather}
where $\varphi = \arg \braket ab$.

For example,  in the vicinity of the fully polarized spin coherent state ($\ket{\phi_0} = \ket{F, m_F^x=F}$), the transverse components of the total angular momentum operator, $\hat{\mathbf J} = \sum_{j=1}^N \hat{\mathbf F}^{(j)}$, define the quadratures of the same atomic oscillator.  The two vectors in Eqs~\eqref {eq:a-vector} and~\eqref {eq:b-vector},
\begin{align}
  \label{eq:y-CSS}
  \ket y &\equiv
  \hat F_y^{(1)} \ket{\phi_0}
  = \sqrt{F/2} \; \ket{F, m_F^x=F-1},
  \\
  \label{eq:z-CSS}
  \ket z &\equiv
  F_z^{(1)} \ket{\phi_0}
  = i \ket y
\end{align}
are parallel to each other with $\varphi = \pi/2$, so $\hat X_y = \hat X_1$ and $\hat P_z = \hat P_1$, and
\begin{gather}
  \label{eq:Jy-Jz-CSS}
  \hat J_y \approx \sqrt{NF} \hat X_1,
  \qquad
  \hat J_z \approx \sqrt{NF} \hat P_1.
\end{gather}

In Appendix~\ref{app:two-oscillators}, we show in general that $\hat X_A$ and $\hat P_B$ defined in Eqs~\eqref {eq:X_A-implicit-def} and~\eqref {eq:X_B-implicit-def} are conjugate quadratures of the same atomic oscillator if and only if $\ket{\phi_0}$ is a minimum uncertainty state with respect to $\hat A^{(1)}$ and $\hat B^{(1)}$.

\paragraph{Orthogonal case.}
If $\ket a$ and $\ket b$ are orthogonal to each other ($\braket ab=0$), then $\hat X_A$ and $\hat P_B$ belong to completely different atomic oscillators.  Choosing $\ket{\phi_1} = \ket a / \|a\|$ and $\ket{\phi_2} = -i \ket b / \|b\|$ yields
\begin{gather}
  \label{eq:Xab-orthogonal}
  \hat X_A = \hat X_1,
  \qquad
  \hat P_B = \hat P_2.
\end{gather}

In recent experiments \cite{PhysRevLett.101.073601, PhysRevLett.99.163002}, the uncertainty in $\hat J_z$ was significantly reduced by coherently squeezing the spin of each individual spin-4 cesium atom via two-axis counter-twisting~\cite{PhysRevA.47.5138} described by the Hamiltonian,
\begin{gather}
  \label{eq:twist}
  \hat T \equiv \sum_{j=1}^N  \frac1{2i}
  \big( \hat F_{x+}^{(j)2} - \hat F_{x-}^{(j)2} \big)
  = \sum_{j=1}^N \big\{ \hat F_y^{(j)}, \hat F_z^{(j)} \big\}.
\end{gather}
Thus each atom is prepared in the same state
\begin{gather}
  \label{eq:twisted-state}
  \ket{\phi_0^{\text{sq}}}
  = \sum_{k=0}^F c_k \ket{F, m_F^x = F - 2k},
\end{gather}
with only even number of spin excitations.  In such a reference state, the quadrature operators assigned to the twisting operator $\hat T$ and to the transverse angular momentum $\hat J_z$ belong to different atomic oscillators.  Following the prescription above, we introduce the corresponding vectors
\begin{align}
  \label{eq:t-vector}
  \ket t &\equiv
  \big(\hat T^{(1)} - \langle T^{(1)} \rangle_0 \big)
  \ket{\phi_0^{\text{sq}}}
  = \hat T^{(1)} \ket{\phi_0^{\text{sq}}},\\
  \label{eq:z-vector}
  \ket z &\equiv
  \big(\hat F_z^{(1)} - \langle F_z^{(1)} \rangle_0 \big)
  \ket{\phi_0^{\text{sq}}}
  = \hat F_z^{(1)} \ket{\phi_0^{\text{sq}}}.
\end{align}
The expectation value $\langle T^{(1)} \rangle_0$ is zero, since it is zero in the initial $x$-polarized spin state, and $\hat T^{(1)}$ is a constant of motion in the process of internal squeezing.  It is easy to see that $\langle F_z^{(1)} \rangle_0$ is also zero.  Noting that $\hat T^{(1)}$ changes the azimuthal quantum number of the hyperfine sublevel by two, whereas $\hat F_z^{(1)}$ changes the same quantum number by only one, we immediately see that $\ket t$ is orthogonal to $\ket z$.  Therefore, completely different atomic oscillators are accessed by the transverse components of the total angular momentum and the twisting operator.

When the reference state is the $x$-polarized coherent spin state and we take the standard spin projection eigenbasis $\ket{\phi_\alpha}=\ket{F,m_F^x=F-\alpha}$, the oscillators associated with $\hat J_z$ and $\hat T$ are the first and second ones, respectively.  The idea of using both these oscillators in a quantum memory and a proposal on how to experimentally access the latter by means of stimulated four-photon processes is presented in Ref.~\cite{PhysRevLett.95.053602}.

We remark here that any process that acts coherently on each atom (e.g., coupling to a classical field) changes the reference state itself but does not influence the atomic oscillators around the reference state.  In this sense, internal spin squeezing as a coherent process does not squeeze any of the collective atomic oscillators.  In particular, it does not squeeze the quadrature operator $\hat P_z$ that is assigned to the $z$ component of the angular momentum.  It does reduce the noise in $\hat J_z$, but only because $\hat J_z \approx [2N (\Delta F_z^{(1)})_0^2]^{1/2} \hat P_z$ and the normalization factor in the brackets is reduced in the new reference state.  In Sec.~\ref{sec:spsq}, we will investigate spin squeezing in more details.

\paragraph{General case.}
If $\ket a$ and $\ket b$ are linearly independent but not orthogonal to each other, then $\hat A$ and $\hat B$ act on different but not independent oscillators.  We can define an orthogonal basis via Gram-Schmidt orthogonalization and we can thus define $\hat X_A = \hat X_1$
and
\begin{gather}
  \label{eq:Xb-general}
  \hat P_B =
  \big( \hat X_1 \cos\varphi + \hat P_1 \sin\varphi \big)
  \cos\vartheta + \hat P_2 \sin\vartheta,
\end{gather}
where $\varphi = \arg \braket ab$ as previously and $\cos\vartheta = |\braket ab| / ( \|a\| \|b\|)$ describes how parallel the two vectors are.  The quadrature operators in this case have a non-canonical commutation relation, $[\hat X_A, \hat P_B] /i < 1$.  See Appendix~\ref {app:two-oscillators} for a derivation of the formulae.

\subsection{Motion in rotating frame}
\label{sec:rot-frame}

In a typical experiment, the atomic spins are placed in a homogeneous
magnetic field where they precess coherently and independently of each
other.  The reference state then also precesses, and we will describe the
collective excitations as deviations from this time dependent state.

In general, we can split the Hamiltonian into two parts, $\hat H = 
\hat H_0 + \hat H_1$, one that acts coherently on each particle and
another that describes, e.g., interaction with an external quantum
field.  Using the interaction free single-particle evolution operator, $\hat
U_0^{(1)} (t) = \mathcal T \exp -i \int_0^t \hat H_0^{(1)}$, we can
introduce the time dependent single-particle basis
$\ket{\phi_\alpha(t)} = \hat U_0^{(1)} (t) \ket{\phi_\alpha}$ that
defines a set of rotating atomic oscillators.  Then we can define time
dependent occupation number states analogously to
Eq.~\eqref{eq:N-body-basis}, and the creation and annihilation
operators \eqref{eq:a-step} are all rotating accordingly in the
Schr\"odinger picture, for example,
\begin{gather}
  \label{eq:a_Sch(t)}
  \hat a_\alpha (t) =
  \hat U_0 (t) \, \hat a_\alpha \, \hat U_0^\dag(t).
\end{gather}
The rotating occupation number states satisfy the interaction free
Schr\"odinger equation.  In the absence of $\hat H_1$, the state of
the rotating atomic oscillators and the number of excitations remain
unchanged.

In the Heisenberg picture, the free Hamiltonian disappears from the
equation of motion of the rotating atomic oscillator operators.  For
example, $\hat a_{\alpha H} (t) = \hat U^\dag (t) \hat a_\alpha (t)
\hat U(t)$ satisfies
\begin{gather}
  \label{eq:a-eq-motion}
  \frac d{dt} \hat a_{\alpha H} (t) =
  i \big[ \hat H_{1H}(t), \hat a_{\alpha H} (t) \big].
\end{gather}
To solve Eq.~\eqref{eq:a-eq-motion}, we need to express $\hat H_1$ in
terms of the rotating oscillator variables.  We assume that the
interaction does not bring the system far from the rotating reference
state $\bigotimes^N \ket{\phi_0(t)}$.  Then Eqs~\eqref{eq:O-approx2}
and \eqref{eq:O-approx1} can be used to express the symmetric
collective atomic operators in $\hat H_1$.  Consider, for example,
that $\hat H_1$ is an arbitrary function of the collective operator
$\hat{\mathcal O}$.  Then we write
\begin{multline}
  \label{eq:O-approx-rot}
  \hat{\mathcal O} \approx N \mathcal O_{00}(t)
  + \sqrt N \sum_{\alpha\ne0}
  \bigl[ \mathcal O_{\alpha0}(t) \hat a_\alpha^\dag (t)
  + \mbox{H.c.} \bigr] \\
  + \sum_{\alpha\beta\ne0}
  \big[\mathcal O_{\alpha\beta}(t)
  - \mathcal O_{00}(t) \delta_{\alpha\beta}\big]
  \hat a_\alpha^\dag(t)   \hat a_\beta(t)
\end{multline}
in place of $\hat{\mathcal O}$, where the matrix elements are
\begin{gather}
  \label{eq:O-ab-rot}
  \mathcal O_{\alpha\beta} (t)
  = \bra{\phi_\alpha} \hat U_0^{(1)\dag}(t)
  \hat{\mathcal O}^{(1)} \hat U_0^{(1)}(t) \ket{\phi_\beta}.
\end{gather}
Depending on whether $\ket{\phi_0(t)}$ is an eigenstate of $\hat{\mathcal
  O}^{(1)}$ or not, we may neglect the quadratic terms in Eq.~\eqref
{eq:O-approx-rot}.  Often $\hat H_1$ is at most quadratic
in the oscillator operators, and the equation of motion reduces to a
set of ordinary linear differential equations that can be solved.

\section{Spin squeezing}
\label{sec:spsq}

For ensembles of spin-$\frac12$ particles, spin squeezing necessarily involves correlation among the spins~\cite{Nature.409.63}.  When these particles are actually the fictitious spins constituting the real spin-$F$ atoms (e.g., the valence electron and the nucleons), correlations may be both intra- and interatomic.  In the former case, the atoms are internally squeezed independently of each other, while in the latter case, there is genuine multi-atomic correlation.  In this section, we investigate how the two ways of squeezing are related to each other.

\subsection{Internal spin squeezing}
\label{sec:internal-squeezing}

We start our analysis with the situation in which the quantum uncertainty in the transverse component of each atomic spin, $\hat F_z^{(j)}$, is reduced independently.  No entanglement is created among the atoms in this way.  To exemplify our analysis, we will consider an ensemble of cesium atoms in the hyperfine level $F=4$ of the atomic ground state $6S_{1/2}$ that is illuminated by an intense laser field.  The light-atom interaction is off-resonantly tuned to the $6S_{1/2}$--$6P_{3/2}$ transition, hence photons are not absorbed, and transitions among the hyperfine sublevels as well as the ground-state energy shifts are mediated only by transfer of photons between different polarization components.  Namely, for the $j$th atom interacting with a light field propagating in the $x$ direction, the tensor light shift reads~\cite {PhysRevA.71.032348}
\begin{multline}
 \label{eq:AL-tensorial}
 \hat H^{(j)}_{AL} = g_0 \hat \phi(0) + g_1 \hat s_z(0) \hat F_x^{(j)}
\\
 + g_2 \Bigl(
 - \hat \phi(0) \hat F_x^{(j)2}
 + \hat s_-(0) \hat F_{x+}^{(j)2}
 + \hat s_+(0) \hat F_{x-}^{(j)2}
 \Bigr),
\end{multline}
where the effective coupling constants $g_i$ sum up all the possible contributions from the different hyperfine levels of the relevant excited states and depend on the detuning as well.  The quantum mechanical Stokes vector components describing the polarization state of light are
\begin{gather}
 \nonumber
 \hat s_+ = \hat s_x + i \hat s_y = \hat a_R^\dag \hat a_L,\\
 \nonumber
 \hat s_z = (\hat  a_R^\dag \hat a_R - \hat  a_L^\dag \hat a_L)/2,\\
 \hat \phi = \hat  a_R^\dag \hat a_R + \hat  a_L^\dag \hat a_L,
 \label{eq:Stokes}
\end{gather}
where $\hat{a}_R$ and $\hat{a}_L$ are annihilation operators for right and left circular polarized, spatially localized photons, and they obey the standard commutation relation $[\hat a_i(z), \hat a_{j}(z')] = c \delta_{ij} \delta (z-z')$.

The photon flux $\hat \phi$ commutes with any other operator in Eq.~\eqref{eq:AL-tensorial} and is usually treated as a c-number.  The quadratic $\hat F_x^{(j)2}$ component can be compensated with the second order Zeeman shift,
\begin{gather}
 \label{eq:2nd-Zeeman}
 \hat H_Z^{(j)} = \omega_L \hat F_x^{(j)} + \beta \hat F_x^{(j)2},
\end{gather}
by tuning the magnetic field such that $\beta = g_2 \phi$ \cite{PhysRevLett.99.163002, PhysRevLett.101.073601}.  To meet the two-photon resonance condition, the right and left polarized light components have to oscillate on the two Larmor sidebands.  After summing over all atoms and performing the rotating wave transformation, we have
\begin{gather}
 \label{eq:H-int-sq}
 \hat H = g_1 \hat s_z(0) \hat J_x
 + 2g_2 \bigl[\hat s_x(0) \hat V
 + \hat s_y(0) \hat T \bigr],
\end{gather}
with the Hermitian collective operators \eqref{eq:twist} and 
\begin{gather}
 \label{eq:twist-tilted}
 \hat V \equiv \sum_{j=1}^N  \frac1{2}
 \big( \hat F_{x+}^{(j)2} + \hat F_{x-}^{(j)2} \big)
 = \sum_{j=1}^N \big( \hat F_y^{(j)2} - \hat F_z^{(j)2} \big).
\end{gather}

We start from the $x$-polarized spin coherent state and take the single-particle angular momentum eigenbasis $\ket{\phi_\alpha} = \ket{F, m_F^x =  F-\alpha}$.  The operator $\hat T$ in Eq.~\eqref{eq:H-int-sq} is responsible for squeezing $\hat F_z^{(j)}$ (and anti-squeezing $\hat F_y^{(j)}$) of each atom, and we can select it by setting $\hat s_y = \phi/2$ to have macroscopic expectation value, while the other two light operators, $\hat s_x$ and $\hat s_z$, can be completely neglected.  This  corresponds to equally strong right and left circularly polarized light components, i.e., a strong linearly polarized field (rotating at twice the Larmor frequency in the laboratory frame).

\begin{figure}
 \centering
 \includegraphics[width=.9\hsize]{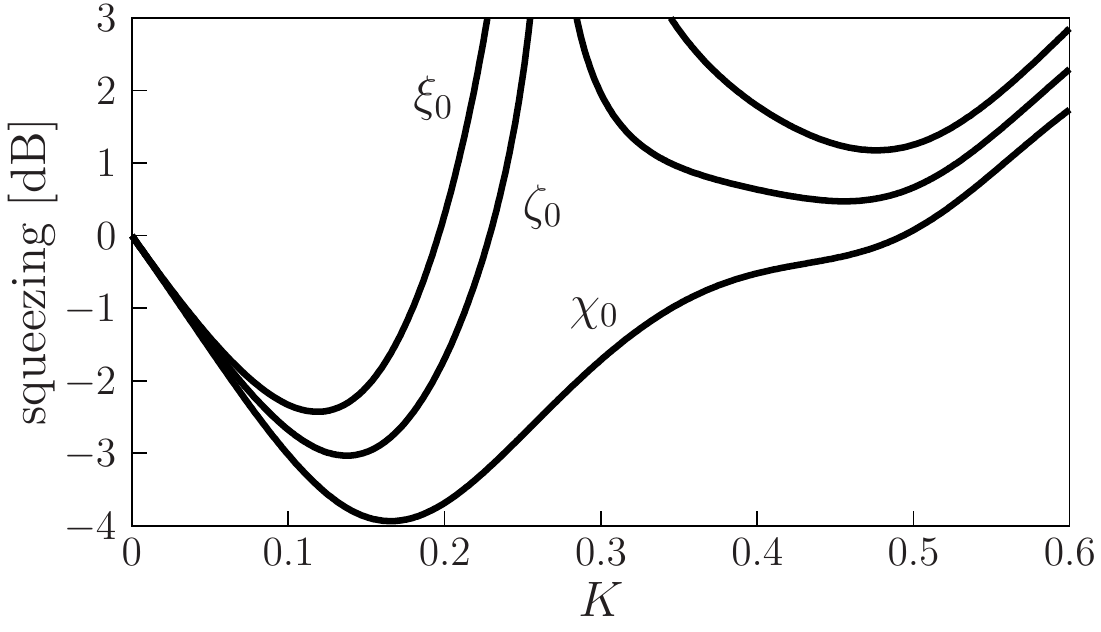}
 \caption{
   Internal spin squeezing, equivalent to squeezing of a single $F=4$
   spin.  The squeezing parameters $\chi_0$, $\zeta_0$, and $\xi_0$
   are shown as function of the integrated interaction strength,
   $K=g_2 \int \phi(t)\,dt$.  The squeezing parameters are, in
   general, defined as follows: $\chi^2 = 2(\Delta J_z)^2 / (NF)$ is
   the uncertainty of the transverse spin component with respect to
   that in the original CSS~\protect \citep{PhysRevA.47.5138},
   $\zeta^2 = 2(\Delta J_z)^2 / \langle J_x \rangle$ is the
   uncertainty compared to that of a CSS with the same longitudinal
   mean spin~\protect \citep{Nature.409.63}, $\xi^2 = 2NF (\Delta
   J_z)^2 / \langle J_x \rangle^2$ is the noise in spin angle~\protect
   \citep{PhysRevA.46.R6797}.  }
 \label{fig:internal-squeezing}
\end{figure}

Under the action of $\hat H_0 = g_2 \phi \hat T$, the initial spin coherent state evolves into the product state $\bigotimes^N [\hat U_0^{(1)} (t) \ket{\phi_0}]$ in which every atom is internally squeezed, while the atomic oscillators expanded around this rotating reference state are in their vacuum states.  Therefore, the statistical properties of collective observables are completely determined by the single-particle expectation values.  In particular, the noise in the transverse component of the total angular momentum is $\Delta J_z = \sqrt N \Delta F_z^{(1)}$, where $\Delta F_z^{(1)}$ is the uncertainty in the transverse spin component of a single atom, and squeezing has the same limit as for a single spin-$F$ particle.  The final degree of squeezing depends only on the dimensionless time integrated interaction strength $K \equiv g_2 \int \phi(t) \, dt$, that is proportional to the total number of photons in the squeezer pulse.  Fig.~\ref{fig:internal-squeezing} shows the degree of squeezing as function of this parameter.

\subsection{Projection based spin squeezing}
\label{sec:qnd-squeezing}

In the next step, we consider a collective process in which the noise in $\hat J_z$ is reduced via quantum non-demolition (QND) measurement and feedback \citep{EurophysLett.42.481, PhysRevLett.85.1594, Science.304.270, PhysRevLett.93.163602, PhysRevLett.93.173002, PhysRevA.65.053819, PhysRevA.65.061801, PhysRevA.70.052324, PhysRevA.72.052313}.  This kind of squeezing creates interatomic entanglement.  To measure $\hat J_z$ in a non-destructive way, we shall couple it to a ``meter'' system, e.g., a light field propagating along the $z$-axis.  This probe field is far detuned, so that we can neglect the second order light shift.  We also neglect spontaneous emission, absorption of the probe beam, and other sources of imperfections that may actually limit spin squeezing \citep{PhysRevA.65.061801, PhysRevA.65.053819, PhysRevA.70.052324, PhysRevLett.93.173002}.  The interaction Hamiltonian in this configuration is $\hat H_1 = g_1 \hat s_z(0) \hat J_z$.  

The Stokes vector component $\hat s_z$ is a QND variable in this interaction: it is not modified by the interaction itself while the light passes different segments of the sample.  It is therefore conventional to treat the accumulated interaction as if the ensemble of atoms as a whole were interacting with a single light mode integrated along the pulse, $\hat S_i (t) \equiv \int_0^{cT} \hat s_i (ct-\xi,t) \,d\xi$.  For a strong classical amplitude populating the $x$-polarized light component, the meter system is the $y$-polarized quantum field and its  quadrature operators can be defined as $\hat X_L \equiv \hat S_y / (c\sqrt{N_p/2})$ and $\hat P_L \equiv \hat S_z / (c\sqrt{N_p/2})$, where $N_p$ is the total photon number in the probe pulse.  Initially, the meter is in the vacuum state and $(\Delta X_L^{\text{in}})^2 = (\Delta P_L^{\text{in}})^2 = \tfrac12$.

Regarding the atomic ensemble, we start from the generic reference product state $\bigotimes^N \ket{\phi_0}$ and assume that the system stays in the vicinity of this state.  We further assume that the polarization in this reference state points in the $x$ direction ($\langle J_y \rangle_0 = \langle J_z \rangle_0 = 0$) and that $\ket{\phi_0}$ is not an eigenstate of $\hat J_z$.  According to our results in Sec.~\ref{sec:linarization}, we assign to $\hat J_z$ a collective atomic oscillator with quadrature variable $\hat P_1 = \hat J_z / {\sqrt{2(\Delta J_z)_0^2}}$. This is the atomic oscillator the light is coupled to and, for the moment, it is enough to consider only this mode.  It is initially in the vacuum state with $(\Delta X_1^{\text{in}})^2 = (\Delta P_1^{\text{in}})^2 = \tfrac12$.  In this generalized Holstein-Primakoff approximation, $\hat P_1$ is also a QND variable. The effective interaction Hamiltonian is $\hat H_1 = g_1 \sqrt{N_p (\Delta J_z)^2_0}\, \hat P_1 \hat P_L /T$, where $T$ is the duration of the probe pulse, and the following input-output relation holds for the quadrature operators in the Heisenberg picture,
\begin{align}
  \label{eq:QND-input-output-1}
  \hat X_1^{\text{out}} &= \hat X_1^{\text{in}} + \kappa \hat P_L^{\text{in}}, &
  \hat P_1^{\text{out}} &= \hat P_1^{\text{in}},\\
  \label{eq:QND-input-output-L}
  \hat X_L^{\text{out}} &= \hat X_L^{\text{in}} + \kappa \hat P_1^{\text{in}}, &
  \hat P_L^{\text{out}} &= \hat P_L^{\text{in}},
\end{align}
where $\kappa \equiv g_1 \sqrt{N_p (\Delta J_z)_0^2}$ is the time integrated coupling that quantifies the strength of the measurement.  The meter system is then read out by measuring $ \hat X_L^{\text{out}}$.  Conditioned on the measurement outcome $x$, the atomic oscillator becomes squeezed.  The new means and variances are \citep{PhysRevA.70.052324}
\begin{align}
  \langle X_1^{\text{out}} \rangle &= 0, &
  (\Delta X_1^{\text{out}})^2 &= (1 + \kappa^2)/2, \\
  \langle P_1^{\text{out}} \rangle &= 
  \frac {x\kappa} {1 + \kappa^2}, &
  (\Delta P_1^{\text{out}})^2 &= 
  \frac12 \frac1{1 + \kappa^2}.
  \label{eq:varP1-qnd}
\end{align}
The uncertainty in $\hat J_z$ thus becomes
\begin{gather}
  \label{eq:varJz-qnd}
  (\Delta J_z^{\text{out}})^2 = \frac{(\Delta J_z)_0^2}{1 + \kappa^2}.
\end{gather}

To ensure that the ensemble is still polarized in the $x$ direction, the measurement result is fed back.  Ideally, the feedback consists of a momentum displacement generated by the position quadrature $\hat X_1$.  In the $x$-polarized coherent spin state or whenever $\ket{\phi_0}$ is a minimum uncertainty state with respect to $\hat F_y^{(1)}$ and $\hat F_z^{(1)}$ (such as the internally squeezed states in Sec.~\ref {sec:internal-squeezing}), the transverse components of the total angular momentum are orthogonal quadratures of the same atomic oscillator, irrespective of the degeneracy of the atomic ground state.  Namely, $\hat J_y$ is proportional to $\hat X_1$, so any Hamiltonian proportional to $\hat J_y$ (e.g., a magnetic field applied in the $y$-axis or a circularly polarized light beam propagating in the $y$ direction) suffices to accommodate the ideal feedback.

For a generic $\ket{\phi_0}$, however, a different feedback operation is required.  The quadrature operator assigned to $\hat J_y$ is now a combination of $\hat X_1$ and $\hat P_1$, and there is also a contribution from an independent second atomic oscillator, as described by Eq.~\eqref {eq:Xb-general}.  Both $\langle J_y^{\text{out}} \rangle$ and $\langle J_z^{\text{out}} \rangle$ are proportional to the measurement outcome, that is, the transverse component of the total angular momentum is slightly tilted in the $yz$ plane.  The feedback apparatus should be appropriately reoriented to correctly cancel this transverse angular momentum component.  We also note that the mean spin $\langle J_x \rangle$ may get further reduced if the feedback procedure excites a second atomic oscillator.

\subsection{Combining the two ways of spin squeezing}
\label{sec:mixed-squeezing}

\begin{figure}
 \centering
 \includegraphics{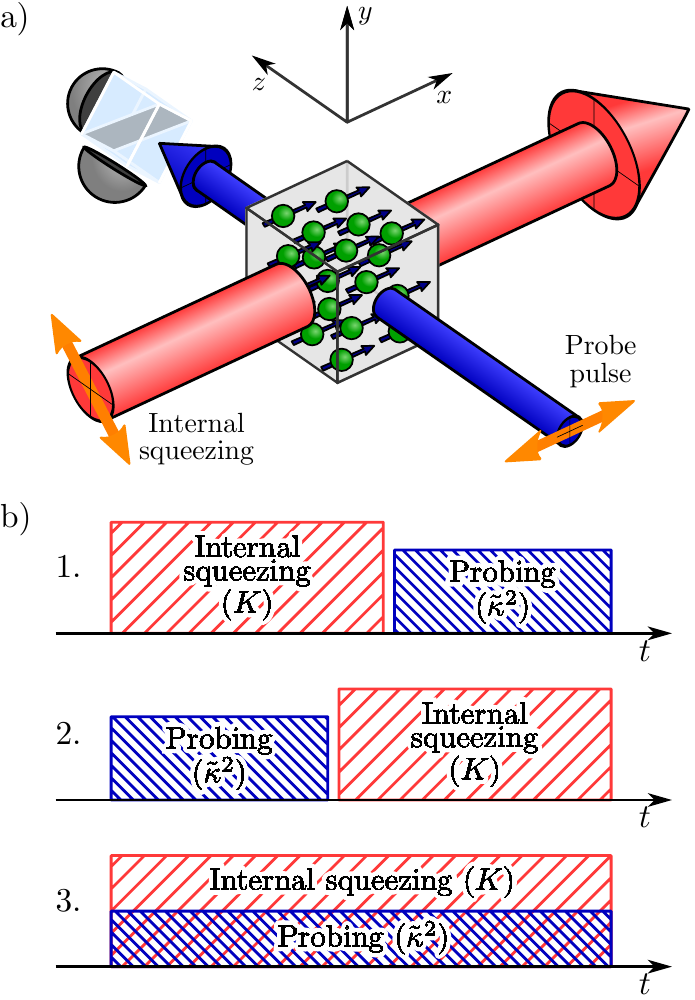}
 \caption{
   (color online) %
   a) Imaginary setup for combining internal squeezing with
   measurement and feedback.  The pulse propagating in the $x$
   direction is responsible for internal squeezing, the other one is
   the probe that measures $\hat J_z$.  b) Pulse diagram of the two
   kinds of squeezing applied in sequence (1,2) and simultaneously (3).
 }
 \label{fig:combi-setup}
\end{figure}

We have seen that spin squeezing can be achieved either by squeezing the corresponding atomic oscillator or by coherently acting on the reference product state of the system.  Here we discuss how the two methods can be combined.  Let us imagine the following oversimplified scenario (Fig.~\ref{fig:combi-setup}).
The atomic ensemble, that is initially prepared in the $x$-polarized coherent spin state, is illuminated (sequentially or simultaneously) by two light pulses:  (i) The squeezer pulse, propagating in the $x$ direction (i.e., in the direction of the atomic polarization), realizes the interaction Hamiltonian $\hat H_0 = g_2 \phi \hat T$ and gives rise to internal squeezing as detailed in Sec.~\ref{sec:internal-squeezing}.  The relevant parameter of this pulse is the time integrated interaction strength $K \equiv g_2 \int \phi(t) \, dt$.  (ii) The probe pulse, propagating in the $z$ direction, has a photon flux $\phi_p(t)$.  The QND interaction with the atomic sample is given by $\hat H_1 = g_1 \hat s_z(0) \hat J_z$.  The Stokes vector component $s_y(z)$ of the outgoing field is continuously measured as in Sec.~\ref{sec:qnd-squeezing}.  The relevant parameter of the probe pulse will be the effective integrated coupling $\tilde \kappa^2 \equiv \frac12 g_1^2 NF \int \phi_p(t) \, dt$.  Given the two pulses, we analyze the following three combinations: (1) internal squeezing is followed by measurement based squeezing, (2) the same sequential squeezing but in reverse order, and (3) when the two methods are applied simultaneously.

\subsubsection{Internal squeezing followed by measurement}
\label{sec:internal+proj}

\begin{figure}
 \centering
 \includegraphics[width=.9\hsize]{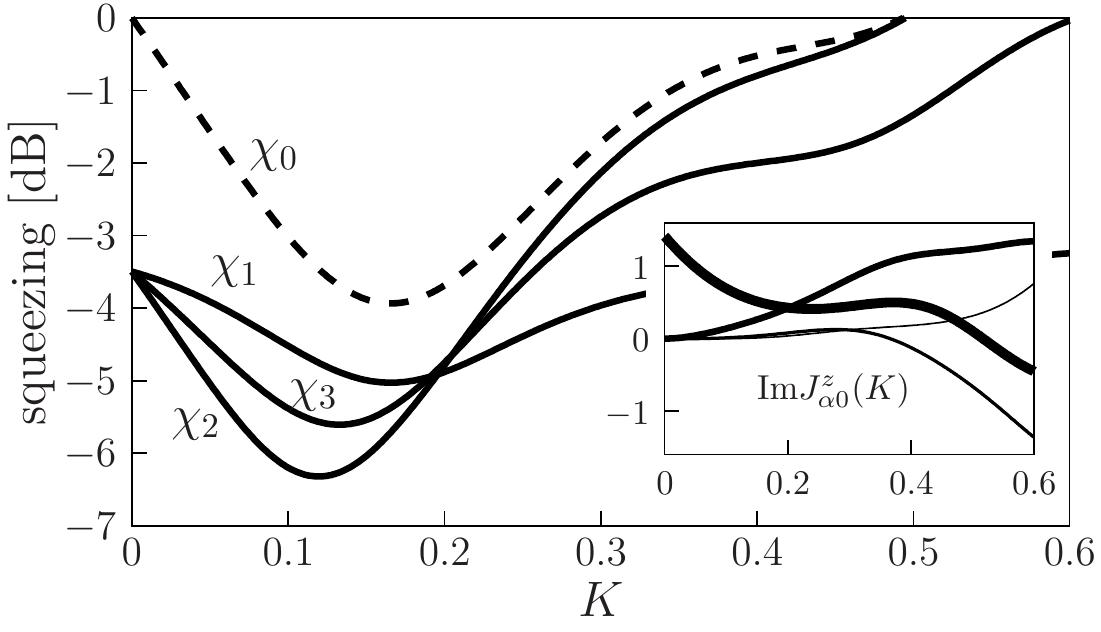}
 \caption{
   Combination of the two ways of spin squeezing, as function of the
   integrated interaction strength of internal squeezing.  Curve
   $\chi_1$ shows the final degree of squeezing when internal
   squeezing is followed by measurement and feedback, while curve
   $\chi_2$ corresponds to the reverse order.  Applying the two
   processes simultaneously results in curve $\chi_3$.  The effective
   time integrated coupling of the measurement is $\tilde\kappa = 2$
   for all the three cases.  As a reference, the dashed line
   ($\chi_0$) shows internal squeezing only.  In the inset: evolution
   of the non-vanishing matrix elements $\Im J^z_{\alpha0}$ for
   $\alpha=1$,~$3$, $5$, and~$7$ (with decreasing line thickness).
 }
 \label{fig:combined-squeezing}
\end{figure}

Let us first address the case in which an internally squeezed ensemble is further squeezed by QND measurement.  In Eq.~\eqref{eq:varJz-qnd}, that already applies for a generic reference state, $(\Delta J_z)_0^2 = \tfrac12 FN$ for the $x$-polarized coherent spin state, but $(\Delta J_z)_0^2 < \tfrac12 FN$ for internally squeezed states.  Therefore, better squeezing can be achieved if the measurement is preceded by internal squeezing.  However, the strength of the measurement also depends on the reference state.  In the same measurement setup (i.e., same probe pulse length and intensity), the time integrated coupling is reduced if the ensemble is internally squeezed: $\kappa = \tilde\kappa \chi_0$, where $\tilde\kappa \equiv g_1 \sqrt{N_p FN/2}$ denotes the coupling for the $x$-polarized coherent spin state, and $\chi_0 = \sqrt{2(\Delta J_z)_0^2 / (NF)}$ is the squeezing parameter of the internally squeezed reference state.  The effective integrated coupling $\tilde\kappa$ does not depend on the actual state of the atomic ensemble, it characterizes only the probe pulse.  The final degree of squeezing, parametrized by $\chi^2 = 2(\Delta J_z)^2/(NF)$, then reads
\begin{gather}
  \label{eq:chi_1}
  \chi_1^2 (K, \tilde \kappa)
  = \big[{\chi_0^{-2}(K) + \tilde\kappa^2}\big]^{-1},
\end{gather}
as function of the integrated interaction strength of the initial internal squeezing process, $K$, and the effective integrated coupling of the measurement, $\tilde \kappa$.  As a comparison, the curve $\chi_0$ in Fig.~\ref {fig:combined-squeezing}
shows internal squeezing only.  The difference between $\chi_1$ and $\chi_0$ is due to the second phase of squeezing, namely, to the QND measurement.  The contribution of the measurement to the overall degree of squeezing decreases with internal squeezing, since the coupling $\kappa = \tilde \kappa \chi_0$ also decreases.

\subsubsection{Measurement and feedback followed by internal squeezing}
\label{sec:proj+internal}

Let us now consider the two squeezing processes in the reverse order.  We start from the $x$-polarized coherent spin state, and consider the single-particle angular momentum eigenbasis $\ket{\phi_\alpha} = \ket{F, m_F^x = F-\alpha}$ as the initial basis for the atomic oscillators.  In this basis, Eq.~\eqref{eq:Jy-Jz-CSS} holds and $\hat P_1$ is the quadrature assigned to $\hat J_z$.  This quadrature is first squeezed by measurement and feedback, so $(\Delta P_1)^2$ becomes as given by Eq.~\eqref{eq:varP1-qnd}, with $\kappa = \tilde \kappa$.  Then we switch off the probe pulse and switch on the squeezer pulse.  Since the Hamiltonian $\hat H_0 = g_2 \phi \hat T$ realizes a coherent action on each atom, the state of the collective atomic oscillators are not changed, only the underlying basis is rotated, $\ket{\phi_\alpha(t)} = \hat U_0(t) \ket{\phi_\alpha}$.  In this new basis, however, $\hat J_z$ no longer corresponds to the squeezed quadrature $\hat P_1$.  Instead, we have
\begin{gather}
  \label{eq:Jz-general}
  \hat J_z \approx \sqrt{2N} \sum_{\alpha\ne0} \Im J_{\alpha0}^z(t) \hat P_\alpha.
\end{gather}
The coefficients $J^z_{\alpha0}(t) = \bra{\phi_\alpha(t)} \hat F_z^{(1)} \ket{\phi_0(t)}$ depend on time only through the integrated interaction strength $K$.  In what follows, we will write $K$ instead of $t$ in the argument.  The functions $J_{\alpha0}^z(K)$ are shown in the inset of Fig.~\ref{fig:combined-squeezing}.  Since the atomic oscillators are independent of each other in this basis, the variance of $\hat J_z$ is given by
\begin{gather}
  \label{eq:DeltaJz-proj+internal}
  \big(\Delta J_z\big)^2 = 2N \sum_{\alpha\ne0} 
  \big[ \Im J_{\alpha0}^z(K) \big]^2 (\Delta P_\alpha^{\text{in}})^2.
\end{gather}
Taking $(\Delta P_1^{\text{in}})^2 = \tfrac12 (1+\tilde\kappa^2)^{-1}$ according to Eq.~\eqref{eq:varP1-qnd}, and $(\Delta P_\alpha^{\text{in}})^2 = \tfrac12$ for $\alpha>1$, the final squeezing parameter is
\begin{gather}
  \label{eq:chi_2}
  \chi_2^2 (K,\tilde\kappa) 
  = \chi_0^2(K) - \frac{[ \Im J_{10}^z(K)]^2 /2}
  {1+\tilde\kappa^{-2}}.
\end{gather}
We can see in Fig.~\ref{fig:combined-squeezing} that the effect of the first, projection based squeezing decreases with the duration of the second, internal squeezing, and the minimum of the overall squeezing is reached before the minimum of $\chi_0$.

\subsubsection{Measuring while squeezing internally}
\label{sec:simultaneous}

Finally we analyze the case when the two kinds of interaction are applied simultaneously.  We assume that the two pulses are switched on at the same time and, for simplicity, we will consider a constant intensity profile for both pulses.  The integrated interaction strength of internal squeezing, $K$, as well as the effective integrated coupling of the measurement, $\tilde\kappa$, can then be controlled via the intensities of the respective pulses.

Since the atomic oscillator basis rotates due to internal squeezing, the probe pulse couples at different time instants to different combinations of the collective atomic oscillators, as specified in Eq.~\eqref{eq:Jz-general}.  Therefore, it is not enough to consider a single atomic oscillator and a single light mode integrated along the pulse.  In Appendix~\ref {app:simultaneous}, we formulate the proper Maxwell-Bloch equations of motion and derive a differential equation for the covariance matrix of the collective atomic quadratures.  Given the time evolution of the reference frame, we can numerically evaluate the squeezing parameter (see $\chi_3$ in Fig.~\ref{fig:combined-squeezing}).

Finally, we note that the presence of an external magnetic field, that is essential for realizing the internal squeezing Hamiltonian, actually prohibits the QND measurement of $\hat J_z$.  This is due to the fact that the transverse angular momentum precesses in the $yz$ plane.  This problem can be circumvented by using two oppositely oriented cells of atoms as in~\cite{Nature.432.482}.

To conclude this section, we have shown that the two kinds of squeezing reduce each others efficiency, and their effect do not simply add up.  If internal squeezing is applied first, the coupling strength of the QND readout is decreased with respect to that in a coherent spin state.  If, on the other hand, the measurement based squeezing is applied first, internal squeezing will mix the collective atomic oscillators, and it is no longer the originally squeezed oscillator that corresponds to the angular momentum component.

\section{Application to QND memories}
\label{sec:qmem}

In atomic quantum memories based on off-resonant QND interaction \cite{Nature.432.482, PhysRevA.70.044304, PhysRevA.73.022331, PhysRevA.73.062329, PhysRevA.74.011802}, the ensemble is usually prepared in a coherent spin state, and the atoms are effectively treated as spin-$\frac12$ particles, for which the conventional Holstein-Primakoff approximation leads to a straightforward oscillator description. We now generalize this approach to arbitrary atomic level structures, and we investigate the interplay between the choice of reference internal state and the collective variables.

Let us now briefly recall the functioning of the quantum memory based on QND measurement and feedback~\cite {Nature.432.482}, but with emphasis on more general reference states.  We consider the same QND setup as in Sec.~\ref {sec:qnd-squeezing}: the atomic ensemble, that has a mean spin pointing to the $x$ direction but is otherwise prepared in a generic reference state, is illuminated by a strong $x$-polarized light field propagating in the $z$ direction.  The photonic quantum oscillator, whose state we wish to map on the ensemble, is the copropagating $y$-polarized quantum field integrated along the pulse.  Irrespective of what the reference state actually is, the information is stored as a disturbance around the arbitrary reference state, and this disturbance can be described as excitation of the collective atomic oscillator which is assigned to $\hat J_z$ in the given reference state.

After the light has first passed the sample, the input-output relation among the quadrature variables of this collective atomic oscillator and those of the photonic oscillator is readily given by Eqs~\eqref {eq:QND-input-output-1} and~\eqref {eq:QND-input-output-L}.  The $\hat X_L^{\text{out}}$ quadrature of the outgoing light field is then measured, and the measurement outcome is fed back.  The atomic variables after the feedback become \cite {PhysRevA.72.052313, Hammerer08}
\begin{gather}
  \hat X_1'' = \hat X_1^{\text{in}} + \kappa \hat P_L^{\text{in}},
  \quad
  \hat P_1'' = -\frac 1\kappa \hat X_L^{\text{in}},
\end{gather}
and both $\hat X_L^{\text{in}}$ and $\hat P_L^{\text{in}}$ have now been transferred to the collective atomic oscillator variables $\hat P_1$ and $\hat X_1$, respectively.  The transfer of information is, however, not perfect: $\eta^2 \equiv 2(\Delta X_1^{\text{in}})^2/\kappa^2$ units of vacuum noise is added to the original $\hat P_L^{\text{in}}$ quadrature due to the initial uncertainty in the collective atomic quadrature $\hat X_1^{\text{in}}$.  It has been proposed to mitigate this imperfection by squeezing the atomic spin state before the memory operation.  Such squeezing operation could be performed, for example, by an additional QND measurement of $\hat J_y$.  Indeed, this would create interatomic correlations and squeeze the collective atomic oscillator.  It has also been suggested~\cite{PhysRevLett.101.073601} that internal squeezing can enhance the fidelity of the quantum memory as well.  We have shown, however, that internal squeezing does not reduce $\Delta X_1$, so it is not clear whether it really enhances the performance of the memory.

We have seen in Sec.~\ref{sec:internal+proj} that the integrated coupling $\kappa = \tilde \kappa \chi_{z0}$ decreases when $\hat J_z$ is squeezed internally, while it increases if $\hat J_z$ is anti-squeezed, for example, due to internal squeezing of $\hat J_y$ instead of $\hat J_z$.  Assuming for a moment that $\ket{\phi_0}$ is a minimum uncertainty state and $(\Delta J_y)_0 (\Delta J_z)_0 = NF/4$, we obtain for the amount of additional noise
\begin{gather}
  \label{eq:memory-noise}
  \eta^2 = \frac {2(\Delta X_1^{\text{in}})^2}{\kappa^2}
  =  \frac {\chi_y^2}{\tilde \kappa^2},
\end{gather}
where $\chi_y$ is the squeezing parameter for $\hat J_y$ and $\tilde \kappa$ is the effective coupling of the QND mapping.  We thus observe that the fidelity of the quantum memory may improve if $\hat J_y$ is internally squeezed.  This does not happen because the collective atomic oscillator is squeezed, but rather because the reference state $\ket{\phi_0}$ is a minimum uncertainty state, and the adjoint, anti-squeezed, $\hat J_z$ implies an increased coupling $\kappa$.

We remark here that $(\Delta J_z)_0^2$ and, thus, the coupling $\kappa$ can  be increased even without squeezing $\hat J_y$.  The reference state is, however, not a minimum uncertainty state in such a case, so the transverse angular momentum components no longer correspond to the same collective atomic oscillator.  This fact has two consequences.  The first regards the feedback and has already been addressed at the end of Sec.~\ref{sec:qnd-squeezing}: by reorienting the feedback apparatus, the measurement outcome can be eliminated from the state of the first collective atomic oscillator, but at the cost of exciting a second one.  The second consequence concerns the read-out of the memory.

The memory can be read out by a similar procedure, interchanging the role of the atomic and photonic oscillators.  When a read-out light pulse passes the atomic ensemble, the quadrature operators are transformed into
\begin{align}
  \hat X_1''' &= \hat X_1'' + \kappa' \hat P_R^{\text{in}},&
  \hat P_1''' &= \hat P_1'',\\
  \hat X_R''' &= \hat X_R^{\text{in}} + \kappa' \hat P_1'',&
  \hat P_R''' &= \hat P_R^{\text{in}}.
\end{align}
Then a measurement of the atomic $\hat X_1$ quadrature and a subsequent feedback onto the light pulse should follow.  In Ref.~\cite {PhysRevA.73.022331}, an additional light beam was suggested to couple a ``meter'' system to the angular momentum component $\hat J_y$.  If the reference state is not a minimum uncertainty state, however, we may not have access to the $\hat X_1$ quadrature through $\hat J_y$.  In a generic reference state, 
\begin{gather}
  \hat J_y \propto 
  \big( \hat X_1 \cos\varphi + \hat P_1 \sin\varphi \big)
  \cos\vartheta + \hat X_2 \sin\vartheta,
\end{gather}
with $\tan \varphi = - \langle F_x^{(1)} \rangle_0 / \langle T^{(1)} \rangle_0$ and $\cos^2 \vartheta = \big(\langle F_x^{(1)} \rangle_0^2 + \langle T^{(1)} \rangle_0^2 \big) \big/ \big[ 4 (\Delta F_y^{(1)})_0^2 (\Delta F_z^{(1)})_0^2 \big]$. 
The $\hat P_1$ component can be ruled out by reorienting the measurement device in the same way as for the feedback.  The contribution from the second atomic oscillator $\hat X_2$, however, cannot be eliminated.  Instead of measuring $\hat X_1$, the best we can actually measure is $x' = \hat X_1''' + \frac {\tan \vartheta}{\cos \varphi} \hat X_2'''$.  The outgoing light quadrature then turns into $\hat P_R''' = \frac 1{\kappa'}(x' - \hat X_1'' - \frac {\tan \vartheta} {\cos \varphi} \hat X_2'')$, and the quadratures of the read-out pulse after the feedback on the light read
\begin{gather}
  \hat X_R^{\text{out}} = -\frac{\kappa'}\kappa
  \hat X_L^{\text{in}} + \hat X_R^{\text{in}},
  \\
  \hat P_R^{\text{out}} = -\frac{\kappa}{\kappa'} \left(
    \hat P_L^{\text{in}} + \frac 1{\kappa}  \hat X_1^{\text{in}}
    + \frac {\tan\vartheta}{\kappa\cos\varphi} \hat X_2^{\text{in}}
  \right).
\end{gather}
Besides the noise term in Eq.~\eqref{eq:memory-noise} and the noise introduced by the $\hat X_R$ quadrature of the readout pulse, an additional noise term appears if $\ket{\phi_0}$ is not a minimum uncertainty state.  Assuming that both the atomic oscillators are initially in their vacuum states, the amount of noise in the $\hat P_R^{\text{out}}$ quadrature (in vacuum noise units) reads
\begin{gather}
  \eta^2 
  = \frac{4 (\Delta F_y^{(1)})_0^2 (\Delta F_z^{(1)})_0^2 
    - \langle F_x^{(1)} \rangle_0^2}{\kappa^2 \langle T^{(1)} \rangle_0^2}
  \ge \frac1{\tilde\kappa^2 \chi_{z0}^2},
\end{gather}
where the lower bound is obtained using the Cauchy-Schwarz inequality~\eqref{eq:cauchy-schwarz}.

To summarize this section, we have analyzed the operation of the quantum memory based on QND interaction and feedback in a generic reference state.  We have shown that, although internal squeezing does not squeeze the collective atomic oscillators, it may enhance the performance of the memory by enhancing the coupling strength of the QND interaction.  We have also pointed out the difficulties arising when the reference state is not a minimum uncertainty state with respect to the transverse angular momentum components.

\section{Conclusions}
\label{sec:conclusions}

If each spin-$\tfrac12$ particle in an ensemble is prepared in the same single-particle state, it is always a spin coherent state, and small perturbations to the product state can be well described by a single collective oscillator degree of freedom in the Holstein-Primakoff approximation.  In this paper, we have introduced a generalization of this method to describe an ensemble of $d$-level atoms in the vicinity of an arbitrary product state (not necessarily spin coherent state).  We have defined $d-1$ independent collective atomic oscillator modes, and we have specified how to express collective operators (namely, permutation invariant sums of single-particle operators) in terms of the oscillator creation and annihilation operators.

We have applied our formalism in particular to spin squeezing of atoms.  We have analyzed two different methods: internal squeezing and QND measurement based squeezing, and identified the collective oscillators that have become squeezed.  We have shown that the two kinds of squeezing reduce the effect of each other.  When the atoms are first internally squeezed, the coupling strength of the QND measurement is reduced and, therefore, the measurement based squeezing is not so efficient.  If we first project the ensemble's state and then continue with internal squeezing, then the latter process will mix the collective atomic oscillators and, at the end of the process, it is no longer the originally squeezed oscillator that corresponds to the transverse angular momentum component.  We have also considered the case when the two kinds of squeezing is applied simultaneously.

Finally, we have analyzed a quantum memory scheme for storing quantum states of light in atomic ensembles based on QND interaction and feedback using an arbitrary initial product state.  We have shown that internal squeezing can reduce the noise of the memory as an indirect effect because of an enhanced coupling strength, caused in fact by the anti-squeezing of the adjoint atomic spin component.


\appendix
\section{}
\label{app:two-oscillators}

In this appendix, we derive the formulae in Sec.~\ref{sec:overlap}.
We define the single-particle basis $\ket{\phi_1}$ and $\ket{\phi_2}$
from the vectors $\ket a$ and $\ket b$ using Gram-Schmidt
orthogonalization, $\ket{\phi_1} = \ket a / \|a\|$ and
\begin{gather}
  \label{eq:gram-schmidt}
  \ket{\phi_2} = -i \frac{ \ket b - (\braket ab /\|a\|^2) \ket a}
  {\sqrt{\|b\|^2 - |\braket ab|^2 / \|a\|^2 }}.
\end{gather}
The matrix elements in Eq.~\eqref{eq:O-approx1b} are
\begin{align}
  A_{10} &= \|a\|,
  &
  B_{10} &= \braket ab / \|a\|,
  \nonumber
  \\
  A_{20} &= 0,
  &
  B_{20} &= i \sqrt{\|b\|^2 - |\braket ab|^2 / \|a\|^2}.
\end{align}
Introducing the mixing angles $\varphi = \arg \braket ab$ and
$\cos\vartheta = |\braket ab| / ( \|a\| \|b\|)$, we can write
\begin{gather}
  \label{eq:B_210}
  B_{10} = \|b\| e^{i\varphi} \cos\vartheta,
  \qquad
  B_{20} = i \|b\| \sin\vartheta.
\end{gather}
Given that $\sqrt{2(\Delta B)_0^2} = \sqrt{2N} \|b\|$, after substituting Eq.~\eqref{eq:B_210} into \eqref{eq:O-approx1b} and comparing it to Eq.~\eqref{eq:X_B-implicit-def}, we arrive at Eq.~\eqref{eq:Xb-general}, which we wanted to prove.

Now we show that it is sufficient and necessary for the quadrature operators $\hat X_A$ and $\hat P_B$ to belong to the same atomic oscillator and to be conjugate to each other that $\ket{\phi_0}$ is a minimum uncertainty state with respect to the single-particle operators $\hat A^{(1)}$ and $\hat B^{(1)}$.  From Eq.~\eqref{eq:Xb-general} and from the definition of the mixing angles, we have the commutation relation
\begin{gather}
  \label{XP-commute-non-canon}
  [\hat X_A, \hat P_B] /i 
  = \frac{\Im \braket ab}{\|a\| \|b\|}
  = \frac{\tfrac1{2i} \langle [ \hat A^{(1)}, \hat B^{(1)} ] \rangle_0}
  {(\Delta A^{(1)})_0 (\Delta B^{(1)})_0}.
\end{gather}
From the Heisenberg uncertainty relation we know that the absolute value of the real number at the right-hand side of Eq.~\eqref{XP-commute-non-canon} is less than or equal to 1, and the inequality is saturated, by definition, for minimum uncertainty states.  Exactly for such states $[\hat X_A, \hat P_B] = \pm i$, and the two quadratures are canonically conjugate to each other.  The Cauchy-Schwarz inequality,
\begin{gather}
  \label{eq:cauchy-schwarz}
  \|a\| \|b\| 
  \ge \sqrt{ (\Re \braket ab)^2 + (\Im \braket ab)^2 }
  ,
\end{gather}
then implies that $\ket a$ and $\ket b$ are parallel to each other ($\cos\vartheta=0$) and $\braket ab$ is pure imaginary ($\varphi=\pm\tfrac\pi2$) if and only if $\ket{\phi_0}$ is a minimum uncertainty state.

\section{}
\label{app:simultaneous}

Here we derive the degree of squeezing when measurement based and internal squeezing is simultaneously applied. See Fig.~\ref{fig:combi-setup} and Sec.~\ref{sec:mixed-squeezing} for a description the setup.

Internal squeezing, governed by the Hamiltonian $\hat H_0 = g_2 \phi \hat T$, acts coherently on each atom.  The corresponding time dependent single-particle basis $\ket{\phi_\alpha(t)} = \hat U_0^{(1)} (t) \ket{\phi_\alpha}$ defines the rotating atomic oscillators.  The coupling to the meter system is described by the interaction Hamiltonian $\hat H_1 = g_1 \hat s_z(0) \hat J_z$, where $\hat s_z$ now refers to the probe pulse, whose photon flux is $\hat\phi_p(z,t) = \phi_p (t-z/c)$.  With the coordinate change $\xi \equiv ct-z$, it is convenient to consider the propagating slices of the $y$-polarized quantum field of the meter system.  The quadrature operators of the slice, that enters the sample at the time instance $\xi/c$, are
\begin{gather}
  \label{eq:xL-pL-stokes}
  \hat x_L(\xi,t) \equiv
  \frac{\hat s_y(ct-\xi,t)}{\sqrt{\phi_p(\xi/c)/2}},
  \;\;\;
  \hat p_L(\xi,t) \equiv 
  \frac{\hat s_z(ct-\xi,t)}{\sqrt{\phi_p(\xi/c)/2}},
\end{gather}
in the Heisenberg picture.  The equation of motion for the Heisenberg operators of the rotating atomic oscillators is given by Eq.~\eqref{eq:a-eq-motion}.  First we express the interaction Hamiltonian with the quadrature operators using Eqs~\eqref {eq:Jz-general} and~\eqref {eq:xL-pL-stokes}, 
\begin{gather}
  \hat H_1(t) \approx g_1 \sqrt{N\phi_p(t)} \sum_{\alpha\ne0}
  \Im J_{\alpha0}^z (t) \, \hat P_\alpha (t) \, \hat p_L (ct,t).
\end{gather}
Then we arrive at the following Maxwell-Bloch equations:
\begin{gather}
  \nonumber
  \frac d{dt} \hat P_\alpha (t) = 0,\qquad
  \frac \partial{\partial t} \hat p_L (\xi,t) = 0,\\
  \nonumber
  \frac d{dt} \hat X_\alpha (t) = g_1 \sqrt{N\phi_p(t)} \,
  \Im J_{\alpha0}^z (t) \, \hat p_L (ct,t),\\
  \nonumber 
  \frac \partial{\partial t} \hat x_L (\xi,t)
  = g_1 \sqrt{N\phi_p(t)}
  \sum_{\alpha\ne0} \Im J_{\alpha0}^z (t) \,
  \hat P_\alpha (t)\, c \delta(ct-\xi).
\end{gather}

To analyze the effect of the continuous light measurement on the atomic oscillators, let us divide the probe pulse into short segments of duration $\tau$.  The quadrature operators of the segment, that enters the sample at the time instance $\xi/c$, are $\hat X_\xi(t) \equiv \int_\xi^{\xi+c\tau} \hat x_L(\xi',t) \, d\xi' /(c\sqrt\tau)$, and similarly for $\hat P_\xi(t)$.  We assume that the change in the reference state due to internal squeezing, as well as the change in the probe photon flux $\phi_p(t)$ can be neglected during the passage of a single light segment, and that the evolution of the atomic system can be obtained by sequential interaction with the segments.  Right after the segment has passed the sample, the following input-output relation holds,
\begin{gather}
  \label{eq:IO1}
  \hat P_\alpha^{\text{out}} = \hat P_\alpha^{\text{in}},\quad
  \hat P_\xi^{\text{out}} = \hat P_\xi^{\text{in}},\\
  \label{eq:IO2}
  \hat X_\alpha^{\text{out}} = \hat X_\alpha^{\text{in}}
  + \kappa_\xi\, \Im J_{\alpha0}^z(\xi/c) \,   \hat P_\xi^{\text{in}},\\
  \label{eq:IO3}
  \hat X_\xi^{\text{out}} = \hat X_\xi^{\text{in}}
  + \kappa_\xi \sum_{\alpha\ne0} \Im J_{\alpha0}^z(\xi/c) \,
  \hat P_\alpha^{\text{in}},
\end{gather}
where the labels ``in'' and ``out'' respectively mean before and after the passage of the light segment in consideration, and $\kappa_\xi \equiv g_1 \sqrt{\phi_p(\xi/c)\tau N}$ is an effective coupling constant.  Introducing the vector $\hat{\mathbf y} \equiv (\hat X_1, \hat P_1, \ldots, \hat X_{2F}, \hat P_{2F}, \hat X_\xi, \hat P_\xi)^T$, we can write Eqs~\eqref{eq:IO1}--\eqref{eq:IO3} as a matrix equation $\hat{\mathbf y}^{\text{out}} = \mathbf S \hat{\mathbf y}^{\text{in}}$.  For Gaussian states, as in the case considered here, the system is fully characterized by the vector of expectation values $\langle \mathbf y \rangle$ and the covariance matrix $\Gamma_{ij} \equiv \big\langle (y_i-\langle y_i\rangle) (y_j-\langle y_j\rangle) \big\rangle$, that transform as $\langle \mathbf y^{\text{out}} \rangle = \mathbf S \langle \mathbf y^{\text{in}} \rangle$ and $\mathbf \Gamma^{\text{out}} = \mathbf S \mathbf \Gamma^{\text{in}} \mathbf S^T$, respectively.

When the light segment enters the sample, it is completely uncorrelated with the atomic ensemble: the incoming covariance matrix is block diagonal, $\mathbf \Gamma^{\text{in}} = \diag (\mathbf \Gamma_{\text{at}}^{\text{in}}, \openone)$, where $\mathbf \Gamma_{\text{at}}$ is the covariance matrix of the collective atomic oscillators and $\openone$ is the $2\times2$ identity matrix describing the initial vacuum state of the light segment.  We denote the outgoing covariance matrix by
\begin{gather}
  \mathbf \Gamma^{\text{out}} =
  \begin{pmatrix}
      \mathbf \Gamma_{\text{at}}^{\text{out}}& \mathbf C\\
      \mathbf C^T& \mathbf \Gamma_\xi^{\text{out}}
  \end{pmatrix},
\end{gather}
where $\mathbf C$ is a $2\times2F$ matrix describing the light-atom correlations.  After the interaction, the meter system is measured.  Conditioned on the measurement outcome, the effect of the light measurement on the atomic covariance matrix is given by the relation~\cite{IntJQuantInf.1.479}
\begin{gather}
  \mathbf \Gamma_{\text{at}}' =
  \mathbf \Gamma_{\text{at}}^{\text{in}}
  - \frac1{[\Gamma_\xi^{\text{out}}]_{11}} \mathbf C 
  \begin{pmatrix} 1&0\\0&0 \end{pmatrix} \mathbf C^T.
\end{gather}

Combining the formulas above, we can write a difference equation that describes the change in the atomic covariance matrix due to the weak QND measurement~\cite{PhysRevA.70.052324}.  If the segment is short enough, it is sufficient to keep only the leading order in $\kappa_\xi^2$, and in the limit of infinitesimal $\tau$, we arrive at a set of ordinary differential equations for the atomic covariance matrix $\mathbf \Gamma$.  In our case, we have a closed subset of equations for the matrix elements describing momentum-momentum correlations, $\gamma_{\alpha\beta} \equiv \big \langle (P_\alpha - \langle P_\alpha \rangle) (P_\beta - \langle P_\beta \rangle) \big \rangle$,
\begin{gather}
  \label{eq:gamma-diffeq}
  \frac d{dt} \gamma_{\alpha\beta} = -\frac{\kappa_\xi^2}\tau
  \sum_{\alpha' \beta'} \gamma_{\alpha\alpha'} \, \Im J_{\alpha'0}^z
  \, \Im J_{\beta'0}^z \, \gamma_{\beta'\beta},
\end{gather}
where $\kappa_\xi^2 / \tau = g_1^2 N \phi_p$.  If the time dependent coefficients $\Im J_{\alpha0}^z (t)$ are known, Eq.~\eqref {eq:gamma-diffeq} can be solved.  Then we can calculate the uncertainty in the transverse angular momentum component \eqref{eq:Jz-general} and the squeezing parameter
\begin{gather}
  \label{eq:chi3}
  \chi_3^2 = \frac12 \sum_{\alpha\beta} \Im J_{\alpha0}^z
  \, \Im J_{\beta0}^z \, \gamma_{\alpha\beta}.
\end{gather}
Fig.~\ref{fig:combined-squeezing} shows the final degree of squeezing obtained by numerically integrating Eq.~\eqref{eq:gamma-diffeq} for constant intensity profiles for both the squeezer pulse and the probe pulse.

\bibliographystyle{myfullaps}
\bibliography{qmem-refs}

\begin{thebibliography}{23}
\providecommand*\natexlab[1]{#1}
\providecommand*\bibnamefont[1]{#1}
\providecommand*\bibfnamefont[1]{#1}
\providecommand*\citenamefont[1]{#1}
\providecommand*\url[1]{\texttt{#1}}
\providecommand*\urlprefix{URL }
\providecommand*{\bibinfo}[2]{#2}
\providecommand*{\eprint}[2][]{\url{#2}}
\providecommand*{\citenameetal}{\textit{et~al.}}
\hbadness=3000

\bibitem[{\citenamefont{Chaudhury \citenameetal}(2007)\citenamefont{Chaudhury,
  Merkel, Herr, Silberfarb, Deutsch, , and Jessen}}]{PhysRevLett.99.163002}
\bibinfo{author}{\bibfnamefont{S.}~\bibnamefont{Chaudhury}},
  \bibinfo{author}{\bibfnamefont{S.}~\bibnamefont{Merkel}},
  \bibinfo{author}{\bibfnamefont{T.}~\bibnamefont{Herr}},
  \bibinfo{author}{\bibfnamefont{A.}~\bibnamefont{Silberfarb}},
  \bibinfo{author}{\bibfnamefont{I.~H.} \bibnamefont{Deutsch}}, ,
  \bibnamefont{and} \bibinfo{author}{\bibfnamefont{P.~S.}
  \bibnamefont{Jessen}}, \emph{\bibinfo{title}{Quantum control of the hyperfine
  spin of a {C}s atom ensemble}}, \bibinfo{journal}{Phys. Rev. Lett.}
  \textbf{\bibinfo{volume}{99}}, \bibinfo{pages}{163002}
  (\bibinfo{year}{2007}).

\bibitem[{\citenamefont{Fernholz \citenameetal}(2008)\citenamefont{Fernholz,
  Krauter, Jensen, Sherson, S{\o}rensen, and Polzik}}]{PhysRevLett.101.073601}
\bibinfo{author}{\bibfnamefont{T.}~\bibnamefont{Fernholz}},
  \bibinfo{author}{\bibfnamefont{H.}~\bibnamefont{Krauter}},
  \bibinfo{author}{\bibfnamefont{K.}~\bibnamefont{Jensen}},
  \bibinfo{author}{\bibfnamefont{J.~F.} \bibnamefont{Sherson}},
  \bibinfo{author}{\bibfnamefont{A.~S.} \bibnamefont{S{\o}rensen}},
  \bibnamefont{and} \bibinfo{author}{\bibfnamefont{E.~S.}
  \bibnamefont{Polzik}}, \emph{\bibinfo{title}{Spin squeezing of atomic
  ensembles via nuclear-electronic spin entanglement}}, \bibinfo{journal}{Phys.
  Rev. Lett.} \textbf{\bibinfo{volume}{101}}, \bibinfo{pages}{073601}
  (\bibinfo{year}{2008}).

\bibitem[{\citenamefont{Kitagawa and Ueda}(1993)}]{PhysRevA.47.5138}
\bibinfo{author}{\bibfnamefont{M.}~\bibnamefont{Kitagawa}} \bibnamefont{and}
  \bibinfo{author}{\bibfnamefont{M.}~\bibnamefont{Ueda}},
  \emph{\bibinfo{title}{Squeezed spin states}}, \bibinfo{journal}{Phys. Rev. A}
  \textbf{\bibinfo{volume}{47}}, \bibinfo{pages}{5138--5143}
  (\bibinfo{year}{1993}).

\bibitem[{\citenamefont{Opatrn{\'y} and
  Fiur\'a\v{s}ek}(2005)}]{PhysRevLett.95.053602}
\bibinfo{author}{\bibfnamefont{T.}~\bibnamefont{Opatrn{\'y}}} \bibnamefont{and}
  \bibinfo{author}{\bibfnamefont{J.}~\bibnamefont{Fiur\'a\v{s}ek}},
  \emph{\bibinfo{title}{Enhancing the capacity and performance of collective
  atomic quantum memory}}, \bibinfo{journal}{Phys. Rev. Lett.}
  \textbf{\bibinfo{volume}{95}}, \bibinfo{pages}{053602}
  (\bibinfo{year}{2005}).

\bibitem[{\citenamefont{S{\o}rensen
  \citenameetal}(2002)\citenamefont{S{\o}rensen, Duan, Cirac, and
  Zoller}}]{Nature.409.63}
\bibinfo{author}{\bibfnamefont{A.}~\bibnamefont{S{\o}rensen}},
  \bibinfo{author}{\bibfnamefont{L.-M.} \bibnamefont{Duan}},
  \bibinfo{author}{\bibfnamefont{J.~I.} \bibnamefont{Cirac}}, \bibnamefont{and}
  \bibinfo{author}{\bibfnamefont{P.}~\bibnamefont{Zoller}},
  \emph{\bibinfo{title}{Many-particle entanglement with bose-einstein
  condensates}}, \bibinfo{journal}{Nature} \textbf{\bibinfo{volume}{409}},
  \bibinfo{pages}{63--66} (\bibinfo{year}{2002}).

\bibitem[{\citenamefont{Kupriyanov
  \citenameetal}(2005)\citenamefont{Kupriyanov, Mishina, Sokolov, Julsgaard,
  and Polzik}}]{PhysRevA.71.032348}
\bibinfo{author}{\bibfnamefont{D.~V.} \bibnamefont{Kupriyanov}},
  \bibinfo{author}{\bibfnamefont{O.~S.} \bibnamefont{Mishina}},
  \bibinfo{author}{\bibfnamefont{I.~M.} \bibnamefont{Sokolov}},
  \bibinfo{author}{\bibfnamefont{B.}~\bibnamefont{Julsgaard}},
  \bibnamefont{and} \bibinfo{author}{\bibfnamefont{E.~S.}
  \bibnamefont{Polzik}}, \emph{\bibinfo{title}{Multimode entanglement of light
  and atomic ensembles via off-resonant coherent forward scattering}},
  \bibinfo{journal}{Phys. Rev. A} \textbf{\bibinfo{volume}{71}},
  \bibinfo{pages}{032348} (\bibinfo{year}{2005}).

\bibitem[{\citenamefont{Wineland \citenameetal}(1992)\citenamefont{Wineland,
  Bollinger, Itano, Moore, and Heinzen}}]{PhysRevA.46.R6797}
\bibinfo{author}{\bibfnamefont{D.~J.} \bibnamefont{Wineland}},
  \bibinfo{author}{\bibfnamefont{J.~J.} \bibnamefont{Bollinger}},
  \bibinfo{author}{\bibfnamefont{W.~M.} \bibnamefont{Itano}},
  \bibinfo{author}{\bibfnamefont{F.~L.} \bibnamefont{Moore}}, \bibnamefont{and}
  \bibinfo{author}{\bibfnamefont{D.~J.} \bibnamefont{Heinzen}},
  \emph{\bibinfo{title}{Spin squeezing and reduced quantum noise in
  spectroscopy}}, \bibinfo{journal}{Phys. Rev. A}
  \textbf{\bibinfo{volume}{46}}, \bibinfo{pages}{R6797--R6800}
  (\bibinfo{year}{1992}).

\bibitem[{\citenamefont{Kuzmich \citenameetal}(1998)\citenamefont{Kuzmich,
  Bigelow, and Mandel}}]{EurophysLett.42.481}
\bibinfo{author}{\bibfnamefont{A.}~\bibnamefont{Kuzmich}},
  \bibinfo{author}{\bibfnamefont{N.~P.} \bibnamefont{Bigelow}},
  \bibnamefont{and} \bibinfo{author}{\bibfnamefont{L.}~\bibnamefont{Mandel}},
  \emph{\bibinfo{title}{Atomic quantum non-demolition measurements and
  squeezing}}, \bibinfo{journal}{Europhys. Lett.}
  \textbf{\bibinfo{volume}{42}}, \bibinfo{pages}{481--486}
  (\bibinfo{year}{1998}).

\bibitem[{\citenamefont{Kuzmich \citenameetal}(2000)\citenamefont{Kuzmich,
  Mandel, and Bigelow}}]{PhysRevLett.85.1594}
\bibinfo{author}{\bibfnamefont{A.}~\bibnamefont{Kuzmich}},
  \bibinfo{author}{\bibfnamefont{L.}~\bibnamefont{Mandel}}, \bibnamefont{and}
  \bibinfo{author}{\bibfnamefont{N.~P.} \bibnamefont{Bigelow}},
  \emph{\bibinfo{title}{Generation of spin squeezing via continuous quantum
  nondemolition measurement}}, \bibinfo{journal}{Phys. Rev. Lett.}
  \textbf{\bibinfo{volume}{85}}, \bibinfo{pages}{1594--1597}
  (\bibinfo{year}{2000}).

\bibitem[{\citenamefont{Geremia \citenameetal}(2004)\citenamefont{Geremia,
  Stockton, and Mabuchi}}]{Science.304.270}
\bibinfo{author}{\bibfnamefont{J.}~\bibnamefont{Geremia}},
  \bibinfo{author}{\bibfnamefont{J.~K.} \bibnamefont{Stockton}},
  \bibnamefont{and} \bibinfo{author}{\bibfnamefont{H.}~\bibnamefont{Mabuchi}},
  \emph{\bibinfo{title}{Real-time quantum feedback control of atomic
  spin-squeezing}}, \bibinfo{journal}{Science} \textbf{\bibinfo{volume}{304}},
  \bibinfo{pages}{270--273} (\bibinfo{year}{2004}).

\bibitem[{\citenamefont{Smith \citenameetal}(2004)\citenamefont{Smith,
  Chaudhury, Silberfarb, Deutsch, and Jessen}}]{PhysRevLett.93.163602}
\bibinfo{author}{\bibfnamefont{G.~A.} \bibnamefont{Smith}},
  \bibinfo{author}{\bibfnamefont{S.}~\bibnamefont{Chaudhury}},
  \bibinfo{author}{\bibfnamefont{A.}~\bibnamefont{Silberfarb}},
  \bibinfo{author}{\bibfnamefont{I.~H.} \bibnamefont{Deutsch}},
  \bibnamefont{and} \bibinfo{author}{\bibfnamefont{P.~S.}
  \bibnamefont{Jessen}}, \emph{\bibinfo{title}{Continuous weak measurement and
  nonlinear dynamics in a cold spin ensemble}}, \bibinfo{journal}{Phys. Rev.
  Lett.} \textbf{\bibinfo{volume}{93}}, \bibinfo{pages}{163602}
  (\bibinfo{year}{2004}).

\bibitem[{\citenamefont{Auzinsh \citenameetal}(2004)\citenamefont{Auzinsh,
  Budker, Kimball, Rochester, Stalnaker, Sushkov, and
  Yashchuk}}]{PhysRevLett.93.173002}
\bibinfo{author}{\bibfnamefont{M.}~\bibnamefont{Auzinsh}},
  \bibinfo{author}{\bibfnamefont{D.}~\bibnamefont{Budker}},
  \bibinfo{author}{\bibfnamefont{D.~F.} \bibnamefont{Kimball}},
  \bibinfo{author}{\bibfnamefont{S.~M.} \bibnamefont{Rochester}},
  \bibinfo{author}{\bibfnamefont{J.~E.} \bibnamefont{Stalnaker}},
  \bibinfo{author}{\bibfnamefont{A.~O.} \bibnamefont{Sushkov}},
  \bibnamefont{and} \bibinfo{author}{\bibfnamefont{V.~V.}
  \bibnamefont{Yashchuk}}, \emph{\bibinfo{title}{Can a quantum nondemolition
  measurement improve the sensitivity of an atomic magnetometer?}},
  \bibinfo{journal}{Phys. Rev. Lett.} \textbf{\bibinfo{volume}{93}},
  \bibinfo{pages}{173002} (\bibinfo{year}{2004}).

\bibitem[{\citenamefont{Andre and Lukin}(2002)}]{PhysRevA.65.053819}
\bibinfo{author}{\bibfnamefont{A.}~\bibnamefont{Andre}} \bibnamefont{and}
  \bibinfo{author}{\bibfnamefont{M.~D.} \bibnamefont{Lukin}},
  \emph{\bibinfo{title}{Atom correlations and spin squeezing near the
  {H}eisenberg limit: Finite-size effect and decoherence}},
  \bibinfo{journal}{Phys. Rev. A} \textbf{\bibinfo{volume}{65}},
  \bibinfo{pages}{053819} (\bibinfo{year}{2002}).

\bibitem[{\citenamefont{Thomsen \citenameetal}(2002)\citenamefont{Thomsen,
  Mancini, and Wiseman}}]{PhysRevA.65.061801}
\bibinfo{author}{\bibfnamefont{L.~K.} \bibnamefont{Thomsen}},
  \bibinfo{author}{\bibfnamefont{S.}~\bibnamefont{Mancini}}, \bibnamefont{and}
  \bibinfo{author}{\bibfnamefont{H.~M.} \bibnamefont{Wiseman}},
  \emph{\bibinfo{title}{Spin squeezing via quantum feedback}},
  \bibinfo{journal}{Phys. Rev. A} \textbf{\bibinfo{volume}{65}},
  \bibinfo{pages}{061801} (\bibinfo{year}{2002}).

\bibitem[{\citenamefont{Madsen and M{\o}lmer}(2004)}]{PhysRevA.70.052324}
\bibinfo{author}{\bibfnamefont{L.~B.} \bibnamefont{Madsen}} \bibnamefont{and}
  \bibinfo{author}{\bibfnamefont{K.}~\bibnamefont{M{\o}lmer}},
  \emph{\bibinfo{title}{Spin squeezing and precision probing with light and
  samples of atoms in the gaussian description}}, \bibinfo{journal}{Phys. Rev.
  A} \textbf{\bibinfo{volume}{70}}, \bibinfo{pages}{052324}
  (\bibinfo{year}{2004}).

\bibitem[{\citenamefont{Hammerer \citenameetal}(2005)\citenamefont{Hammerer,
  Polzik, and Cirac}}]{PhysRevA.72.052313}
\bibinfo{author}{\bibfnamefont{K.}~\bibnamefont{Hammerer}},
  \bibinfo{author}{\bibfnamefont{E.~S.} \bibnamefont{Polzik}},
  \bibnamefont{and} \bibinfo{author}{\bibfnamefont{J.~I.} \bibnamefont{Cirac}},
  \emph{\bibinfo{title}{Teleportation and spin squeezing utilizing multimode
  entanglement of light with atoms}}, \bibinfo{journal}{Phys. Rev. A}
  \textbf{\bibinfo{volume}{72}}, \bibinfo{pages}{052313}
  (\bibinfo{year}{2005}).

\bibitem[{\citenamefont{Julsgaard \citenameetal}(2004)\citenamefont{Julsgaard,
  Sherson, Cirac, Fiur\'a\v{s}ek, and Polzik}}]{Nature.432.482}
\bibinfo{author}{\bibfnamefont{B.}~\bibnamefont{Julsgaard}},
  \bibinfo{author}{\bibfnamefont{J.}~\bibnamefont{Sherson}},
  \bibinfo{author}{\bibfnamefont{J.~I.} \bibnamefont{Cirac}},
  \bibinfo{author}{\bibfnamefont{J.}~\bibnamefont{Fiur\'a\v{s}ek}},
  \bibnamefont{and} \bibinfo{author}{\bibfnamefont{E.~S.}
  \bibnamefont{Polzik}}, \emph{\bibinfo{title}{Experimental demonstration of
  quantum memory for light}}, \bibinfo{journal}{Nature}
  \textbf{\bibinfo{volume}{432}}, \bibinfo{pages}{482--486}
  (\bibinfo{year}{2004}).

\bibitem[{\citenamefont{Hammerer \citenameetal}(2004)\citenamefont{Hammerer,
  M{\o}lmer, Polzik, and Cirac}}]{PhysRevA.70.044304}
\bibinfo{author}{\bibfnamefont{K.}~\bibnamefont{Hammerer}},
  \bibinfo{author}{\bibfnamefont{K.}~\bibnamefont{M{\o}lmer}},
  \bibinfo{author}{\bibfnamefont{E.~S.} \bibnamefont{Polzik}},
  \bibnamefont{and} \bibinfo{author}{\bibfnamefont{J.~I.} \bibnamefont{Cirac}},
  \emph{\bibinfo{title}{Light-matter quantum interface}},
  \bibinfo{journal}{Phys. Rev. A} \textbf{\bibinfo{volume}{70}},
  \bibinfo{pages}{044304} (\bibinfo{year}{2004}).

\bibitem[{\citenamefont{Fiur\'a\v{s}ek
  \citenameetal}(2006)\citenamefont{Fiur\'a\v{s}ek, Sherson, Opatrn{\'y}, and
  Polzik}}]{PhysRevA.73.022331}
\bibinfo{author}{\bibfnamefont{J.}~\bibnamefont{Fiur\'a\v{s}ek}},
  \bibinfo{author}{\bibfnamefont{J.}~\bibnamefont{Sherson}},
  \bibinfo{author}{\bibfnamefont{T.}~\bibnamefont{Opatrn{\'y}}},
  \bibnamefont{and} \bibinfo{author}{\bibfnamefont{E.~S.}
  \bibnamefont{Polzik}}, \emph{\bibinfo{title}{Single-passage readout of atomic
  quantum memory}}, \bibinfo{journal}{Phys. Rev. A}
  \textbf{\bibinfo{volume}{73}}, \bibinfo{pages}{022331}
  (\bibinfo{year}{2006}).

\bibitem[{\citenamefont{Muschik \citenameetal}(2006)\citenamefont{Muschik,
  Hammerer, Polzik, and Cirac}}]{PhysRevA.73.062329}
\bibinfo{author}{\bibfnamefont{C.~A.} \bibnamefont{Muschik}},
  \bibinfo{author}{\bibfnamefont{K.}~\bibnamefont{Hammerer}},
  \bibinfo{author}{\bibfnamefont{E.~S.} \bibnamefont{Polzik}},
  \bibnamefont{and} \bibinfo{author}{\bibfnamefont{J.~I.} \bibnamefont{Cirac}},
  \emph{\bibinfo{title}{Efficient quantum memory and entanglement between light
  and an atomic ensemble using magnetic fields}}, \bibinfo{journal}{Phys. Rev.
  A} \textbf{\bibinfo{volume}{73}}, \bibinfo{pages}{062329}
  (\bibinfo{year}{2006}).

\bibitem[{\citenamefont{Sherson \citenameetal}(2006)\citenamefont{Sherson,
  S{\o}rensen, Fiur\'a\v{s}ek, M{\o}lmer, and Polzik}}]{PhysRevA.74.011802}
\bibinfo{author}{\bibfnamefont{J.}~\bibnamefont{Sherson}},
  \bibinfo{author}{\bibfnamefont{A.~S.} \bibnamefont{S{\o}rensen}},
  \bibinfo{author}{\bibfnamefont{J.}~\bibnamefont{Fiur\'a\v{s}ek}},
  \bibinfo{author}{\bibfnamefont{K.}~\bibnamefont{M{\o}lmer}},
  \bibnamefont{and} \bibinfo{author}{\bibfnamefont{E.~S.}
  \bibnamefont{Polzik}}, \emph{\bibinfo{title}{Light qubit storage and
  retrieval using macroscopic atomic ensembles}}, \bibinfo{journal}{Phys. Rev.
  A} \textbf{\bibinfo{volume}{74}}, \bibinfo{pages}{011802(R)}
  (\bibinfo{year}{2006}).

\bibitem[{\citenamefont{Hammerer \citenameetal}(2009)\citenamefont{Hammerer,
  S{\o}rensen, and Polzik}}]{Hammerer08}
\bibinfo{author}{\bibfnamefont{K.}~\bibnamefont{Hammerer}},
  \bibinfo{author}{\bibfnamefont{A.~S.} \bibnamefont{S{\o}rensen}},
  \bibnamefont{and} \bibinfo{author}{\bibfnamefont{E.~S.}
  \bibnamefont{Polzik}}, \emph{\bibinfo{title}{Quantum interface between light
  and atomic ensembles}} (\bibinfo{year}{2009}), \eprint{arXiv:0807.3358v4
  [quant-ph]}.

\bibitem[{\citenamefont{Eisert and Plenio}(2003)}]{IntJQuantInf.1.479}
\bibinfo{author}{\bibfnamefont{J.}~\bibnamefont{Eisert}} \bibnamefont{and}
  \bibinfo{author}{\bibfnamefont{M.~B.} \bibnamefont{Plenio}},
  \emph{\bibinfo{title}{Introduction to the basics of entanglement theory in
  continuous-variable systems}}, \bibinfo{journal}{Int. J. Quant. Inf.}
  \textbf{\bibinfo{volume}{1}}, \bibinfo{pages}{479--506}
  (\bibinfo{year}{2003}).

\end{thebibliography}
\end{document}